\documentclass[printer]{aa}
\usepackage{graphicx}
\usepackage{txfonts}
\usepackage{natbib}
\usepackage{longtable}
\bibpunct{(}{)}{;}{a}{}{,}

\begin{document}
\title{On the fundamental line of galactic and extragalactic globular clusters}
\author{M. Pasquato \inst{1}
\and G. Bertin \inst{2}} \institute{Dipartimento di Fisica,
Universit\`a di Pisa, Largo Bruno Pontecorvo 3, I-56127 Pisa, Italy
\and Dipartimento di Fisica, Universit\`a degli Studi di Milano,
via Celoria 16, I-20133 Milano, Italy}
\date{Received XXX / Accepted YYY}
\authorrunning{Pasquato \& Bertin}
\titlerunning{FL of Galactic and Extragalactic GCs}

\abstract{In a previous paper, we found that globular clusters in our Galaxy lie close to a line in the ($\log{R_e}$, ${\mathit{SB}}_e$, $\log{\sigma}$) parameter space, with a moderate degree of scatter and remarkable axi-symmetry. This implies that a purely photometric scaling law exists, that can be obtained by projecting this line onto the ($\log{R_e}$, ${\mathit{SB}}_e$) plane. These photometric quantities are readily available for large samples of clusters, as opposed to stellar velocity dispersion data.}{We study a sample of $129$ Galactic and extragalactic clusters on this photometric plane in the V-band. We search for a linear relation between ${\mathit{SB}}_e$ and $\log{R_e}$ and study how the scatter around the best-fit relation is influenced by both age and dynamical environment. We interpret our results in terms of testing the evolutionary versus primordial origin of the fundamental line.}{We perform a detailed analysis of surface brightness profiles, which allows us to present a catalogue of structural properties without relying on a given dynamical model.}{We find a linear relation between ${\mathit{SB}}_e$ and $\log{R_e}$, in the form ${\mathit{SB}}_e = (5.25 \pm 0.44) \log{R_e} + (15.58 \pm 0.28)$, where ${\mathit{SB}}_e$ is measured in mag/arcsec$^2$ and $R_e$ in parsec. Both young and old clusters follow the scaling law, which has a scatter of approximately $1$ mag in ${\mathit{SB}}_e$. However, young clusters display more of a scatter and a clear trend in this with age, which old clusters do not. This trend becomes tighter if cluster age is measured in units of the cluster half-light relaxation time. Two-body relaxation therefore plays a major role, together with passive stellar population evolution, in shaping the relation between ${\mathit{SB}}_e$, $\log{R_e}$, and cluster age. We argue that the $\log{R_e}$-${\mathit{SB}}_e$ relation and hence the fundamental line scaling law does not have a primordial origin at cluster formation, but is rather the result of a combination of stellar evolution and collisional dynamical evolution.}{} 

\maketitle

\section{Introduction}
\label{Sect:Introduction}

In a previous paper \citep[][hereafter Paper I]{MyFirstPaper}, we demonstrated that Galactic globular clusters (Galactic GCs) occupy a narrow region around a line in the ($\log{R_e}$, ${\mathit{SB}}_e$, $\log{\sigma}$) parameter space. This fundamental line is located within the fundamental plane of Galactic GCs \citep[see][]{Djorgovski1995} and its existence was noted by \cite{Bellazzini} based on principal component analysis techniques.

\cite{Bellazzini} divided the observed Galactic GCs into two sets of inner GCs and outer GCs with respect to the Sun's orbit, and found that outer GCs have a dimensionality of $1$ in the ($\log{R_c}$,  ${\mathit{SB}}_0$, $\log{\sigma}$) space\footnote{The photometric quantities considered by \cite{Bellazzini} are core/central quantities (i.e., core radius $R_c$ and central surface brightness ${\mathit{SB}}_0$), as opposed to the half-light ones adopted in Paper I and in this paper (i.e., half-light radius $R_e$ and average surface brightness within it ${\mathit{SB}}_e$).}, while the dimensionality rises to $2$ for inner GCs. He argued that inner GCs, which are closer to the Galactic bulge and interact more frequently with the disk, are more dynamically disturbed than outer GCs. Therefore, he suggested that dynamical interactions with the environment are more likely to disrupt than to preserve a fundamental line, which was then speculated to be primordial in origin.

By projecting the fundamental line onto the ($\log{R_e}$, ${\mathit{SB}}_e$) plane, in Paper I we predicted the existence of a purely photometric scaling law in the form of a linear relation between ${\mathit{SB}}_e$ and $\log{R_e}$. This relation can be practically studied for a much larger sample than the fundamental plane, because velocity dispersion data is not required. In particular, bright clusters (both young and old) in the LMC, SMC, and Fornax galaxies may be included in the study, together with Galactic GCs. This allows a comparison to be made between the behaviors in the ($\log{R_e}$, ${\mathit{SB}}_e$) plane, of clusters that have significantly different ages and are associated with different environments.

The choice of using only photometric variables allows us to consider a relatively large sample of clusters with a substantial age spread. We are then able to attack the problem of the origin and evolution of the ${\mathit{SB}}_e$-$\log{R_e}$ scaling law by studying its dependence on cluster age and to look for evidence for or against the picture in which this scaling law originates primordially and diminishes with time because of dynamical evolution.

In the present paper, we follow Paper I in deriving cluster structural parameters in a model-independent way, in particular by avoiding fitting a given dynamical model to observational data. Some minor changes in the model-independent algorithm used are described in Sect.~\ref{Sect:TheData}, where a detailed account of the data that we collect from the literature is also given. We compile a catalogue of fraction-of-light radii, average surface brightness, and absolute magnitudes, as a machine-readable table.
Central cusps in the surface brightness profile have been observed in both old, Galactic GCs \citep[][]{NoyolaGalacticSlopes} and extragalactic clusters \citep[][]{NoyolaExtragalacticSlopes} and are predicted to form after core-collapse by N-body simulations \citep[][]{Tr2009}.
Extended cluster envelopes have also been observed in young clusters in external galaxies \citep[e.g.,][]{LarsenEnvelopes, SchweitzerEnvelopes}.
The simple, single-mass, spherical isotropic \cite{KingModels} models that are usually taken to fit cluster surface brightness profiles have a flat core and sharp tidal cutoff and therefore have difficulties in modeling clusters with these features. The problem of extended envelopes prompted \cite{CatMcLaughlin} to refit a sample of Milky Way, LMC, SMC, and Fornax clusters with King models, power-law models, and \cite{WilsonModels} models, finding that, for a sizeable fraction of the old GC population, King models do not perform more effectively than either \cite{WilsonModels} or power-law models in fitting surface brightness profiles. More generally, the construction of physically-motivated self-consistent models continues to advance, by including features such as tidal flattening \citep[e.g., see][]{GBVarri}, but model-independent approaches remain a valid complement to model-based analyses \citep[see][]{KronMayall}.
While a certain dependence on specific assumptions, such as the details of the extrapolation of the external part of surface brightness profiles, cannot be avoided, the catalogue of model-independent GC structural parameters presented in this paper naturally emerges as a tool useful for avoiding the potentially limiting assumptions required by dynamical models.
The fundamental line and its scatter are presented in Sect.~\ref{Sect:Results}. Discussion and conclusions are given in Sect.~\ref{Sect:DiscussionAndConclusions}.

\section{The data}
\label{Sect:TheData}

\subsection{Distance moduli}
We combined the distance moduli by \cite{RecioBlancoDistances} and \cite{FerraroDistances} for the Galactic GCs in our sample.
\cite{RecioBlancoDistances} and \cite{FerraroDistances} list $72$ and $61$ objects respectively, with an overlap of $34$, which provides a sample of $99$ Galactic GCs with measured distance moduli.
Both the apparent distance moduli in the F555W band in \cite{RecioBlancoDistances} and the true distance moduli in \cite{FerraroDistances} are calibrated on the horizontal-branch standard candle. The distance moduli from \cite{FerraroDistances} are taken from Col.~$8$ of their Table $2$, i.e., are based on the $[M/H]$ metallicity scale. We converted the distance moduli by \cite{RecioBlancoDistances} to true distance moduli using the adopted $E(B-V)$ reddenings (see following subsection) and the relation between reddening and extinction in the F555W band from Table $12$ of \cite{CalibrationF555WReddening}. For the data in the overlap of the \cite{FerraroDistances} and \cite{RecioBlancoDistances} samples, we took the error-weighted average value of the true distance modulus. In Paper I, we adopted quite conservative uncertainties in distance moduli, with a rescaling in the error bars to include systematic errors and an estimate of the error caused by the limited knowledge of the amount of $\alpha$-enhancement. This procedure allowed us to prove that the scatter about the fundamental plane is not caused by observational errors alone, even though conservatively estimated. For the purpose of compiling the catalogue described in Sect.~\ref{SubSect:TheCatalogue}, we consider the errorbars quoted by the authors of the respective measurements and combine them in the standard way.
For the extragalactic GCs in our sample, we adopt the distance to the host galaxy as the cluster's distance, following \cite{CatMcLaughlin}, with the exception of the two clusters NGC $1466$ and NGC $1841$ which, while being listed as part of the LMC GC system by \cite{CatMcLaughlin}, have an independent distance modulus measurement by \cite{RecioBlancoDistances}.
The extragalactic sample of \cite{CatMcLaughlin} comprises $68$ GCs. Thus,
the sample of both Galactic and extragalactic GCs with measured distance moduli comprises $165$ objects.

\subsection{Reddening}
For Galactic GCs, we combined reddening data by \cite{FerraroDistances} and \cite{RecioBlancoDistances}. For reddenings present in both listings, we adopted the error-weighted average. Errors are assumed to be $10\%$ of the adopted reddening value, as in Paper I. The clusters NGC $6304$, NGC $6316$, NGC $6342$, NGC $6356$, NGC $6388$, NGC $6441$, NGC $6539$, NGC $6624$, and NGC $6760$ have their distance modulus measured by \cite{RecioBlancoDistances}, but reddening was listed for them by neither \cite{RecioBlancoDistances} nor \cite{FerraroDistances}. Therefore, for these clusters, we chose to use the reddening values provided by \cite{CatHarris}.

\subsection{Surface brightness profiles: apparent magnitudes and effective radii}
The model-independent procedure based on spline-smoothing that we used for extracting integrated apparent magnitudes and fraction-of-light radii from the surface brightness profiles of GCs does not differ substantially from the one described in Paper I.
Three differences are worth mentioning:
\begin{itemize}
	\item The smoothing parameter of the spline-fitting algorithm is now adjusted interactively by directly examining how much the resulting spline matches the GC surface brightness data.
	\item The extrapolation that is taken past the outermost observed point in the surface brightness profile assumes the projected luminosity density decreases as a power law with projected distance from the GC center. A discussion of this power-law behavior of the outer part of the GC surface brightness profiles is given in Sect.~\ref{Sect:Results}. The power law is truncated at $5 \times r_{ou}$, where $r_{ou}$ is the outermost observed radius. The contribution to the integrated luminosity from the extrapolated profile is usually not influenced significantly by changes in this cutoff. The power-law exponent is determined by fitting the outer points in the surface brightness profile. The exact number of points used for each cluster is again defined interactively. GCs with a contribution to the integrated apparent magnitude from the extrapolated profile that exceeds $0.5$ magnitudes are removed from the sample.
	\item Automated outlier rejection based on comparison with literature data does not take place. Instead, the cluster surface brightness profiles are inspected visually, and as a result those of NGC $2419$, NGC $4372$, NGC $6101$, NGC $6541$, and NGC $6535$ are excluded from the sample due to their poor quality. 
\end{itemize}

A sample of $125$ surface brightness profiles was acquired from \cite{CatTrager} for Galactic GCs, and $68$ surface brightness profiles of extragalactic clusters were taken from \cite{CatMcLaughlin}. We have a combined sample of $145$ GCs with model-independent structural parameters, after removing the clusters with an excessive (as defined above) contribution to the apparent luminosity from the extrapolated profile and the five listed explicitly.

\subsection{Shape of the surface brightness profile at the GC center}
In Paper I, we found an unexpected correlation between residuals to the fundamental plane of Galactic GCs and the central logarithmic slope of the GC surface brightness profile defined and measured by \cite{NoyolaGalacticSlopes}. In the present paper, the slopes measured by \cite{NoyolaExtragalacticSlopes} for extragalactic clusters are combined with those measured by \cite{NoyolaGalacticSlopes}, producing a sample of $68$ GCs with measured slopes.

\subsection{Extrapolation of the surface brightness profile}
Most GC surface brightness profiles can be accurately fitted in their outer parts 
with a power-law luminosity density $I$ per unit area given by
\begin{equation}
 \label{Eq:powerlawreal}
 I \propto R^{-a}
\end{equation}

\noindent We determine the value of the exponent $a$ for the sample of $145$ GCs with quality surface brightness profiles for which we compute the model-independent structural parameters.
This best-fit relation is then used to extrapolate the surface brightness profile and calculate the cluster integrated apparent magnitude.

\subsection{Distance from the center of the host galaxy and orbital parameters}
To assess the role of environment-driven dynamical evolution in establishing or eliminating the ${\mathit{SB}}_e$-$\log R_e$ relation, we use distances from 
the center of the host galaxy listed by \cite{CatMcLaughlin} as a quantitative measure of the interaction. This approach is similar to that of \cite{Bellazzini}, who divided Galactic GCs on the basis of Galactocentric distance to study the fundamental line of GCs.
 
In principle, the size of the Roche lobe is a more accurate estimator of the amount of tidal disturbance that the host galaxy exerts on a cluster. A cluster of mass $M_{GC}$ in a circular orbit with an angular velocity $\Omega$ at distance $R_G$ from the center of its host galaxy (taken to be spherically symmetric) has a tidal radius \citep[see][]{GBVarri}
\begin{equation}
 \label{HillSphere}
 r_T = {\left(\frac{G M_{GC}}{\Omega^2 \nu}\right)}^{1/3},
\end{equation}
where $\nu \equiv 4 - \kappa^2/\Omega^2$ is a coefficient that depends on the rotation curve $V = \Omega R$ associated with the galaxy. For a flat rotation curve, $\nu = 2$ and
\begin{equation}
 \label{HillSphere2}
 r_T \propto R^{2/3}_G M^{1/3}_{GC},
\end{equation}
where the constant of proportionality varies from one galaxy to another.
The sample of clusters with model-independent structural parameters, host-galaxy distances, and masses estimated by \cite{CatMcLaughlin} consists of $114$ GCs. We use Eq.~(\ref{HillSphere2}) to calculate the tidal radius for $61$ clusters in the Milky Way and $40$ in the LMC; we do not consider either the SMC or Fornax objects because their sample size would be too small. Of course, a thorough analysis would require a discussion of how the available projected data truly constrain the relevant intrinsic quantities which appear in Eqs.~\ref{HillSphere} and \ref{HillSphere2} \citep[see also][]{Spitzer87, HeHu03} but this is beyond the scope of the present paper. \cite{CatMcLaughlin} list only sky-projected distances for LMC, SMC, and Fornax clusters. Full three-dimensional distances are available only for Galactic GCs.

Measuring the strength of the dynamical interaction with the host galaxy using detailed orbital data instead of instantaneous distances would be more accurate, but these data are available only for a much smaller subset of Galactic GCs. \cite{AMPOrbits48GC} calculate the orbits of $48$ GCs in a model Milky Way potential, yielding average pericenter distances and orbital eccentricities.
The sample of clusters with such an orbit determination and model-independent structural parameters consists of $36$ GCs.
\cite{AMPOrbits48GC} calculate orbits based on GC radial velocity and proper motion data, based on either an axisymmetric or a barred model Galactic potential. In the following, we use orbital data from the axisymmetric model, but we checked that our results do not differ significantly if the barred model is assumed instead.

\subsection{Ages and timescales}
\cite{MarinFranchAges} obtain GC relative ages from HST/ACS data acquired by the ACS Survey of Galactic globular clusters, while \cite{DeAngeliAges} list ages from either ground-based photometry or HST snapshot data. For both samples, we adopt the ages based on the \cite{CarrettaGratton} metallicity scale, normalize both sets of relative ages to NGC $104$, and adopt the average value for the combined samples. NGC $6287$ is the oldest cluster in the combined sample, so we set its age to be $13$ Gyr to determine the absolute ages of the combined set of Galactic GCs.
The young clusters of the LMC, SMC, and Fornax are assigned ages following \cite{CatMcLaughlin}. The final sample of Galactic and extragalactic GCs with measured ages contains $131$ objects.

Relaxation times at the half-light radius are taken from \cite{CatMcLaughlin}. Since \cite{CatMcLaughlin} analyze the GC surface brightness profiles in a model-dependent way, we adopt the values based on \cite{KingModels} model fits.
The processes driven by two-body interactions, such as mass loss by means of the evaporation of stars, mass segregation of stars of different masses within the GC, and core collapse, are expected to occur over a timescale of several relaxation times. The ratio of age to relaxation time is thus a measure of the ``dynamical age'', i.e., the degree of relaxation of a given cluster.

\cite{AMPOrbits48GC} use the orbits that they calculate for their sample of $48$ GCs to compute destruction rates for bulge- and disk-shocking\footnote{The \cite{AMPOrbits48GC} destruction rates differ significantly from other literature values \citep[e.g., see][]{Kr2009}. This should be kept in mind when considering our results in Sect.~\ref{galaenv}.}. These are the reciprocals of the natural timescales against which cluster age is compared to quantify the amount of dynamical interaction with the host Galactic environment a that given GC has suffered.

In the following, we denote cluster ages by $\tau_a$, half-light relaxation times by $\tau_r$, disk-shocking destruction rates by $1/\tau_d$, and bulge-shocking destruction rates by $1/\tau_b$.

\subsection{The catalogue}
\label{SubSect:TheCatalogue}
For the adopted sample of $129$ GCs, Table \ref{MeasuredPhotQTable} lists the integrated apparent $V$-band magnitude obtained by using our model-independent profile smoothing method, the adopted reddening and true distance modulus, the log projected half-light radius (in arcsecs), and the value of the exponent $a$ obtained by fitting a power-law to the outermost part of the GC surface brightness profile (see Eq.~(\ref{Eq:powerlawreal})).
Table~\ref{DerivedPhotQTable} lists the average $V$-band surface brightness within the half-light radius ${\mathit{SB}}_e$ (in mag/arcsec$^2$), half-light radius $\log R_e$ (in parsecs), absolute magnitude $M_V$, and residuals in ${\mathit{SB}}_e$ to the fitted ${\mathit{SB}}_e$-$\log R_e$ linear relation.

\begin{figure}
  \resizebox{\hsize}{!}{\includegraphics[angle = 270]{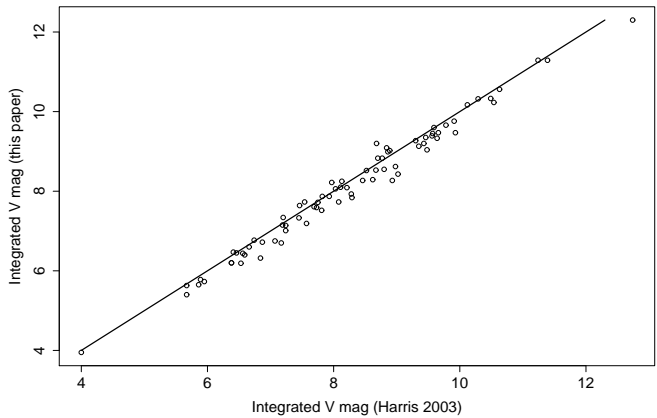}}
  \caption{Comparison between the integrated apparent $V$-band magnitudes of GCs derived in the present paper with the values listed by \cite{CatHarris}. The superimposed line represents the identity. The magnitudes we derive are systematically fainter by approximately $0.12$ mag in the median with respect to the \cite{CatHarris} values.\label{compharrison}}
\end{figure}

To assess the impact of model-dependence on the GC structural parameters, we compare the luminosities that we obtained here with the values listed in the catalogue by \cite{CatHarris} for Galactic GCs. The sample with parameters available both in the Harris catalogue and in this paper contains $76$ GCs. Figure \ref{compharrison} compares our V-band integrated apparent magnitudes with the \cite{CatHarris} values. The magnitudes derived in the present paper are fainter than those of \cite{CatHarris} by $0.12$ mag in the median. The rms scatter of the differences is $0.2$ mag. Nine clusters exhibit a discrepancy of $0.4$ mag or more between the integrated apparent magnitude that we derive here and the \cite{CatHarris} value. These clusters are Arp $2$, NGC $4590$, NGC $5466$, NGC $5634$, NGC $6218$, NGC $6316$, NGC $6366$, NGC $6584$, and NGC $6809$. With the exception of NGC $6584$ and NGC $6316$, a visual inspection of the \cite{CatTrager} surface brightness profiles of these clusters detects either a very noisy profile or puzzling features such as a non-monotonic behavior.

\subsection{Fraction of light radii and non-parametric concentration measurements}
Fraction of light radii for fractions other than $1/2$ are also obtained for each GC and listed in Table \ref{FoLRadiiTable}, thus providing an update and extension of the study by \cite{CatTrager}, who list the fraction of light radii (from $10\%$ to the half-light radius) for their sample of Galactic GCs. We recall that the ratio of two different fraction of light radii defines dimensionless shape parameters for the surface brightness profile, which for King-model clusters should be on a one-to-one relation with the concentration parameter $c$. \cite{CatMcLaughlin} call for ``a more generally applicable concentration index able to represent the spatial extent or potential depth of any cluster in a more model-independent way (so as to allow, e.g., for a combined analysis of clusters that may not all be described well by the same type of model)''.

Using our fraction of light radii, we obtain the coefficients $C_{21} = r_e/r_{25\%}$, $C_{32} = r_{75\%}/r_e$, and $C_{31} = r_{75\%}/r_{25\%}$. These concentration indicators do not rely on model-based assumptions on the shape of the surface brightness profile, and can easily be calculated for a given model surface brightness profile.
We also obtain the so-called Third Galaxy Concentration ($TGC$) index introduced by \cite{NonPaGTC01} to improve the stability of non-parametric indicators of galaxy concentration as a function of the radial extension of the observed luminosity profile. The $TGC$ coefficient is defined to be the ratio of the light contained within $r_e/3$ to the light contained within the half-light radius $r_e$.\footnote{This is the choice that we adopt in the present paper, but in principle fractions other than $1/3$ could be chosen, leading to slightly different definitions of $TGC$.} To the best of our knowledge, this is the first calculation of these indices for GCs, while the use of non-parametric concentration indicators for galaxies dates to the studies of \cite{NonPaF72} and \cite{NonPaD77}.

\begin{figure}
  \resizebox{\hsize}{!}{\includegraphics[angle = 270]{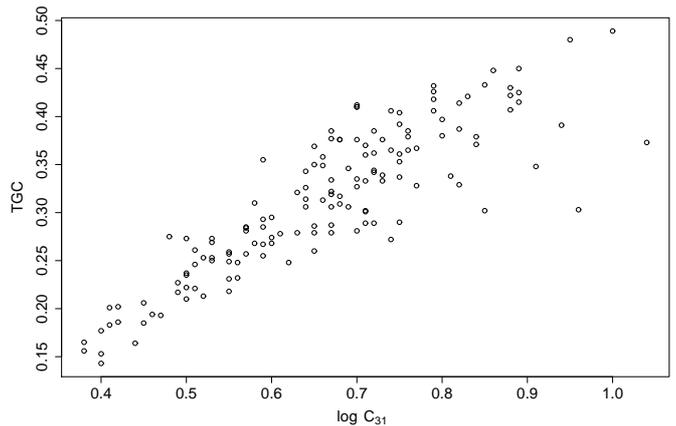}}
  \caption{Relation between the $C_{31}$ and the $TGC$ non-parametric concentration indices for the adopted sample of $145$ GCs with quality surface brightness profiles.  \label{C31TGC}}
\end{figure}

Figure~\ref{C31TGC} shows the $TGC$ index as a function of $\log{C_{31}}$ for the adopted sample of $145$ GCs with quality surface brightness profiles. We find a strong correlation with $r = 0.87$. This is an indirect check that GC surface brightness profiles are described well by a one-parameter family of models. The observed linear correlation still leaves room for ingredients such as mass segregation or IMBHs to introduce deviations from this simple one-parameter picture.

\section{Results}
\label{Sect:Results}
\subsection{The ${\mathit{SB}}_e$-$\log R_e$ relation}
\begin{figure}
  \resizebox{\hsize}{!}{\includegraphics[angle = 270]{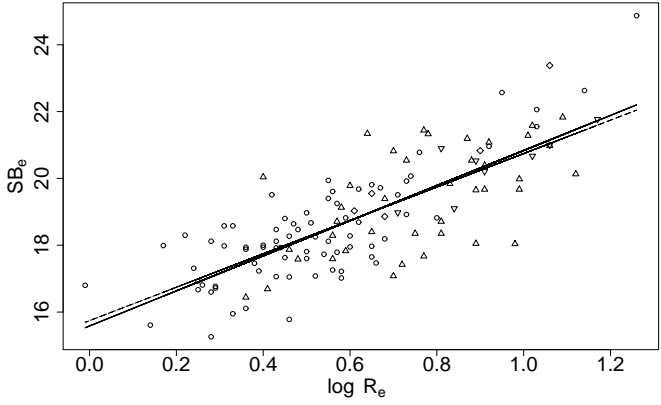}}
  \caption{Relation between V-band ${\mathit{SB}}_e$ and $\log R_e$ for our catalogue of $129$ GCs. Open circles are Galactic GCs, upwards facing triangles are LMC clusters, downwards facing triangles are SMC clusters, and diamonds are Fornax clusters. The solid line is the linear biweight fit of ${\mathit{SB}}_e$ versus $\log R_e$ over the entire sample (see Eq.~\ref{Eq:SBelogReRelat}). The thinner dashed line corresponds to a slope of exactly $5$ (see Eq.~\ref{Eq:KormendiIdentity}). \label{SBelogRe}}
\end{figure}

As anticipated, based on the projected fundamental line argument of Paper I, we do find a linear relation between V-band ${\mathit{SB}}_e$ and $\log R_e$ for the adopted sample of $129$ GCs, of the form
\begin{equation}
\label{Eq:SBelogReRelat}
{\mathit{SB}}_e = (5.25 \pm 0.44) \log{R_e} + (15.58 \pm 0.28),
\end{equation}
where the coefficients are obtained by biweight \citep[e.g., see][]{Biweight} fitting ${\mathit{SB}}_e$ as a function of $\log R_e$, and errors are derived by bootstrap resampling. The scatter about the relation is $1.04$ mag in ${\mathit{SB}}_e$, as quantified by the standard deviation of its residuals.
Figure \ref{SBelogRe} shows the results for our catalogue of $129$ GCs in the ($\log R_e$, ${\mathit{SB}}_e$) plane, different symbols representing the various host galaxies.

Equation (\ref{Eq:SBelogReRelat}) results from neglecting the errors in the individual data-points, an approach which is usually taken in the literature \citep[e.g.,][]{Djorgovski1995}. If, on the other hand, these errors are included in an ordinary least-squares regression fit, we obtain
\begin{equation}
 \label{Eq:IsItSixOrFive}
{\mathit{SB}}_e = (6.09 \pm 0.40) \log{R_e} + (15.11 \pm 0.25).
\end{equation}

The slopes of Eqs.~(\ref{Eq:SBelogReRelat}) and (\ref{Eq:IsItSixOrFive}) are consistent with each other within $3$-$\sigma$.

\subsection{Interpretation}
Equation (\ref{Eq:SBelogReRelat}) is compatible to within $1$ $\sigma$ with
\begin{equation}
 \label{Eq:KormendiIdentity}
 {\mathit{SB}}_e = 5 \log{R_e} + k,
\end{equation}
which, for a given $k$, describes the locus of constant absolute magnitude $M_V = {\mathit{SB}}_e - 5 \log R_e$ within the ($\log R_e$, ${\mathit{SB}}_e$) plane.
Therefore, the clusters of our sample show no systematic trend of absolute magnitude with half-light radius, even though the spanned magnitude range is large.

\begin{figure}
  \resizebox{\hsize}{!}{\includegraphics[angle = 270]{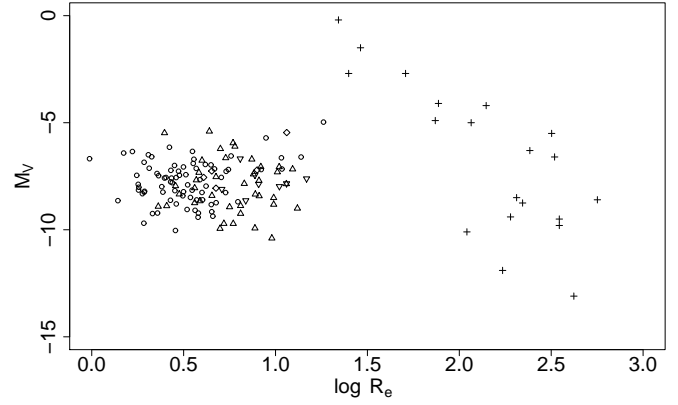}}
  \caption{Absolute integrated V-band magnitude $M_V$ as a function of $\log R_e$ (in parsec) for our sample of $129$ GCs (open circles are Galactic GCs, upwards facing triangles are LMC clusters, downwards facing triangles are SMC clusters, and diamonds are Fornax clusters) and the sample of $21$ nearby dwarf spheroidal galaxies by \cite{vandenBerghNearbyDSPH} (crosses). \label{MvlogRe}}
\end{figure}

This behavior is evident from Fig.~\ref{MvlogRe}, where we plot V-band integrated absolute magnitude $M_V$ as a function of $\log R_e$ for GCs (identified by parent galaxy symbols as in Fig.~\ref{SBelogRe}) and for nearby dwarf spheroidal galaxies (the data for the latter are taken from \citet{vandenBerghNearbyDSPH}). While dwarf spheroidals show a clear trend of absolute magnitude with half-light radius, GCs form an almost perfect scatter plot, independent of their parent galaxy.

Quantitatively, by fitting a linear relation to $M_V$ as a function of $\log R_e$ for the full sample of $129$ GCs, we obtain
\begin{equation}
 \label{MvVSlogRe}
 M_V = (0.3 \pm 0.4) \log R_e - (8.0 \pm 0.2)
\end{equation}
where the $\log R_e$ coefficient does not differ significantly from $0$.

Therefore, the existence of a linear ${\mathit{SB}}_e$-$\log R_e$ relation appears to reflect the statement that GCs have a luminosity independent of size. For a constant mass-to-light ratio $M/L$, this requirement dictates that GCs have a mass distribution that does not depend on their physical size. For Galactic GCs, this is a well-known fact \citep[e.g., see][]{DJMey94}, which has also been recently confirmed for extragalactic clusters \citep[see][]{Barmby} and is at the basis of the explanation proposed by \cite{Bellazzini} for the existence of a fundamental line.

Even if Galactic, LMC, SMC, and Fornax clusters are considered separately, no significant trend of $M_V$ with $\log R_e$ emerges, i.e., the independence of total luminosity on cluster size is apparently universal.

\subsection{A trend of residuals with age for young clusters}

\begin{figure}
  \resizebox{\hsize}{!}{\includegraphics[angle = 270]{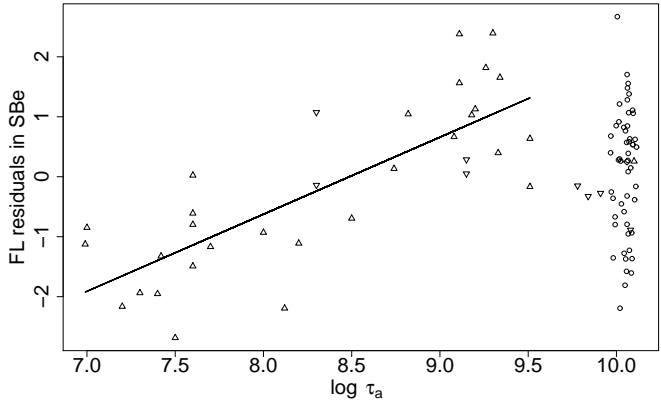}}
  \caption{Trend of residuals to the ${\mathit{SB}}_e$-$\log R_e$ relation with cluster age. Open circles are Galactic GCs, upwards facing triangles are LMC clusters, downwards facing triangles are SMC clusters, and diamonds are Fornax clusters. The solid line is the linear least squares fit of residuals versus log age for clusters younger than $4$ Gyr. The old GCs are not located along the extrapolation of the correlation line for the younger clusters, which indicates that the young, massive clusters that are in the SMC, LMC, and Fornax now, are not likely to be similar to the YMCs that were the progenitors of the old GCs (see text for a discussion). \label{TrendAgeResiduals}}
\end{figure}

\begin{figure}
  \resizebox{\hsize}{!}{\includegraphics[angle = 270]{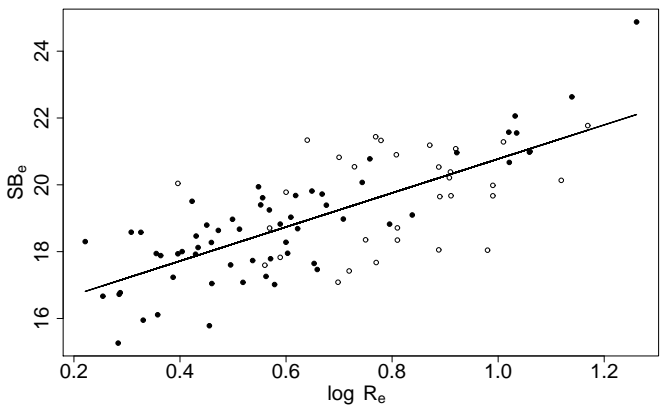}}
  \caption{Relation between V-band ${\mathit{SB}}_e$ and $\log R_e$ for the $91$ GCs in our sample with a reliable age determination. The solid line is the biweight linear fit to this restricted sample, with a slope of $5.09$. Open circles are clusters younger than $4$ Gyr, solid circles are GCs older than $4$ Gyr.\label{SBeReOldYoung}}
\end{figure}

The residuals to the fitted ${\mathit{SB}}_e$-$\log R_e$ relation are plotted against cluster age in Fig.~\ref{TrendAgeResiduals}. The residuals are defined as the difference between the true surface brightness and that predicted by the fitted linear relationship, so that a positive residual means a fainter surface brightness. Based on the assumption of a constant mass-to-light ratio, this implies a lower surface density.
The plot shows a rather different behavior for young and old clusters.
Old clusters, which we define to have age in excess of $4$ Gyr, are mainly Galactic GCs and their residuals show no significant trend with age (the correlation coefficient is $r = 0.08$, compatible with $0$ according to a t-test).
The residuals of clusters younger than $4$ Gyr, instead show a clear trend with age, with $r = 0.79$ and a highly significant t-test p-value of $5.4 \times {10}^{-8}$. Moreover, Fig.~\ref{SBeReOldYoung} shows that these young clusters exhibit a larger scatter about the fitted line, as quantitatively confirmed by the standard deviation of their residuals with respect to it, which is $1.37$ mag in ${\mathit{SB}}_e$ as opposed to $1.04$ mag for the scatter about the relation given by Eq.~(\ref{Eq:SBelogReRelat}) for the whole sample.
The presence of a greater scatter and the indication that it is driven by a third parameter clearly infers that the behavior of young clusters and old clusters in the ($\log R_e$, ${\mathit{SB}}_e$) plane is different.
Moreover, Fig.~\ref{TrendAgeResiduals} shows that old GCs are not located along the extrapolation of the correlation line for the younger clusters. While a firm conclusion regarding this matter would require additional studies, this occurrence may be an indication that massive clusters that are in the SMC, LMC, and Fornax now, are either not similar to the young massive clusters that were the progenitors of the old GCs, or that evolutionary phenomena capable of explaining the non-linearity observed in Fig.~\ref{TrendAgeResiduals} are at work.
 
\subsection{Passive stellar evolution or dynamical aging?}

The trend of the residuals to the fitted ${\mathit{SB}}_e$-$\log R_e$ relation with cluster age may be produced by the evolution of the cluster stellar population with age, whose characteristic timescales are influenced by cluster metallicity but do not display large variations from one cluster to another, or from dynamical effects, e.g., mass loss due to evaporation of light stars. To decouple the dynamical effects from stellar evolution, we studied the dependence of the residuals from the ${\mathit{SB}}_e$-$\log R_e$ relation on the ratio $\tau_a/\tau_r$ of cluster age $\tau_a$ to the half-light relaxation time $\tau_r$.
The timescale for cluster dissolution by the evaporation of stars is a multiple of $\tau_r$, as well as the timescale for the onset of core-collapse and mass segregation. The ratio $\tau_a/\tau_r$ is therefore a measure of \emph{dynamical age}, i.e., of how much a cluster has evolved by two-body relaxation processes.

\begin{figure}
  \resizebox{\hsize}{!}{\includegraphics[angle = 270]{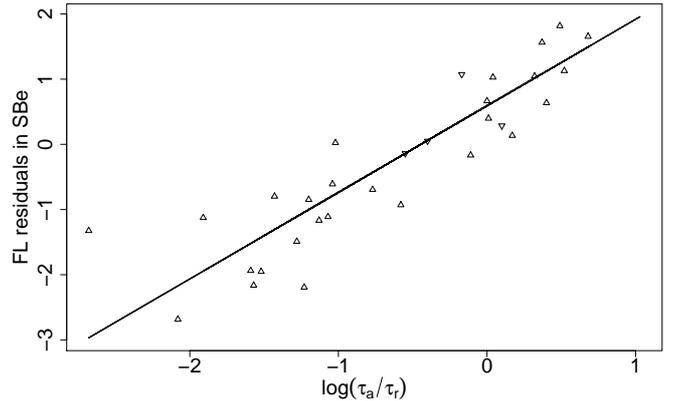}}
  \caption{Trend of residuals to the ${\mathit{SB}}_e$-$\log R_e$ relation with cluster age in units of the cluster half-light relaxation time. Only clusters younger than $4$ Gyr are shown in the plot. Upwards facing triangles are LMC clusters, downwards facing triangles are SMC clusters. The superimposed line is obtained by performing a least squares linear regression.\label{TrendAgeRelaxResiduals}}
\end{figure}

Figure \ref{TrendAgeRelaxResiduals} shows a scatter plot of the residuals against $\tau_a/\tau_r$. Only clusters younger than $4$ Gyr are included in the plot.
The trend shown here is stronger than the correlation of residuals with cluster age alone (not shown). The correlation coefficient for clusters younger than $4$ Gyr is $r = 0.90$, which is quite tight, even though it remains consistent within its $95\%$ confidence interval with the value of $r = 0.79$ found for the correlation with age alone.
This finding suggests that, for young clusters, internal two-body relaxation phenomena play an important role in driving the evolution of the ${\mathit{SB}}_e$-$\log R_e$ relation.

\subsection{Residuals and the central slope of the surface brightness profile}
\begin{figure}
  \resizebox{\hsize}{!}{\includegraphics[angle = 270]{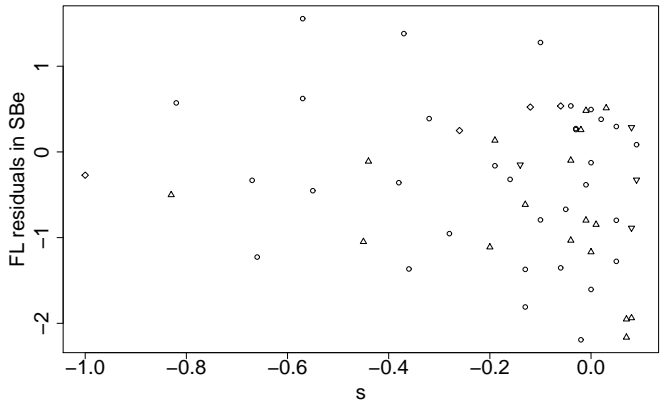}}
  \caption{Scatter plot of the residuals to the ${\mathit{SB}}_e$-$\log R_e$ relation versus \cite{NoyolaGalacticSlopes, NoyolaExtragalacticSlopes} central surface brightness profile slope $s$. Open circles are Galactic GCs, upwards facing triangles are LMC clusters, downwards facing triangles are SMC clusters, and diamonds are Fornax clusters.\label{ScatterPlotSlopes}}
\end{figure}

Figure \ref{ScatterPlotSlopes} shows that there is no apparent correlation between residuals of the ${\mathit{SB}}_e$-$\log R_e$ relation and central surface brightness profile slopes measured by \cite{NoyolaGalacticSlopes, NoyolaExtragalacticSlopes}. The correlation coefficient that we find is $r = -0.19$, which is not significant (with a p-value of $0.16$), i.e., $r$ is compatible with $0$ and no correlation is present. This result is somewhat unexpected, given the correlation found in Paper I between central surface brightness profile slope and residuals to the fundamental plane of GCs. A possible interpretation is that this correlation is caused by phenomena pertaining to the velocity dispersion $\sigma$. This would be the case if the presence of a cusp in the cluster luminosity density profile makes $\sigma$ higher than expected from the fundamental plane relation.

\subsection{Dynamical influence of the galactic environment}
\label{galaenv}
To measure environmental
effects on the ${\mathit{SB}}_e$-$\log R_e$ relation,
we study how the residuals vary with different indicators.
To quantify the amount of influence that the galactic environment exerts on a sample of $114$ GCs, we use distances from the center of each cluster's host galaxy, listed by \cite{CatMcLaughlin}.

We find no significant trend of residuals to the ${\mathit{SB}}_e$-$\log R_e$ relation with log distance from the host galaxy center. A positive correlation coefficient of $r = 0.22$ is found, which corresponds to a t-test p-value of $0.02$. 

For the sample of $61$ Galactic GCs for which we calculated the radii of the relevant Roche lobes, we find a negative correlation coefficient with the residuals of the ${\mathit{SB}}_e$-$\log R_e$ relation, which is marginally significant with $0.01$ p-value. In the case of LMC clusters, no significant trend with tidal radii is found.


For the $48$ Galactic GCs whose orbits were computed by \cite{AMPOrbits48GC}, the interaction with Galactic environment can be quantified in a more refined way, by using the computed orbital parameters. \cite{AMPOrbits48GC} calculated their orbits over a period of several Gyr, and list average values of perigalactic distance and orbital eccentricity during this time. These two orbital parameters are understood to set the cluster tidal cutoff radii \citep[e.g., see][]{EmpiriKing}, thereby influencing the whole cluster density and surface brightness profile.
Surprisingly, we find no correlation between the residuals and either perigalactic distance or orbital eccentricity. The correlation coefficients are consistent with $0$ having p-values of $0.19$ and $0.30$, respectively.

%

\cite{AMPOrbits48GC} also computed GC destruction rates relative to the tidal interaction with either the Galactic bulge or the disk. Destruction rates depend on the computed GC orbit as well as the assumed Galactic potential model. Our results are similar for the two potential models considered by the authors of the above paper. We consider the ratio of GC age to the related timescales to measure the amount of environment-driven dynamical evolution.
A scatter plot of the residuals\footnote{both computed for an axisymmetric Galactic potential model} of the ${\mathit{SB}}_e$-$\log R_e$ relation versus the log ratio of cluster age $\tau_a$ to \cite{AMPOrbits48GC} bulge destruction time $\tau_b$ and disk destruction time $\tau_d$ shows only a mild trend, with correlation coefficients of $0.45$ and $0.49$, respectively.
%
%

\subsection{Power-law behavior of surface brightness profiles in the outermost regions}
\label{Sect:PLbehaviour}

\begin{figure}
  \resizebox{\hsize}{!}{\includegraphics[angle = 270]{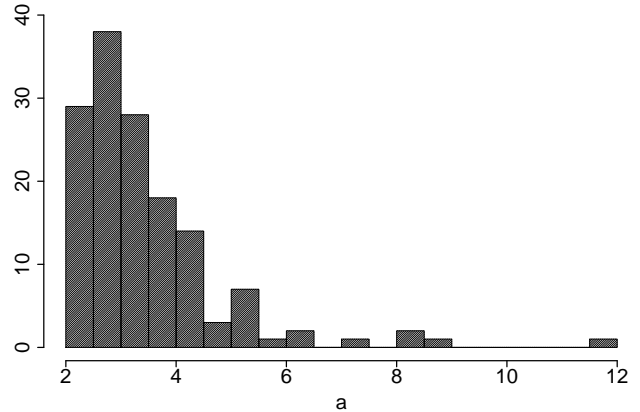}}
  \caption{Histogram of best-fit power law coefficients $a$ for the outer part of the GC surface brightness profiles.\label{HistExponentsPowerLaw}}
\end{figure}

\begin{figure}
  \resizebox{\hsize}{!}{\includegraphics[angle = 270]{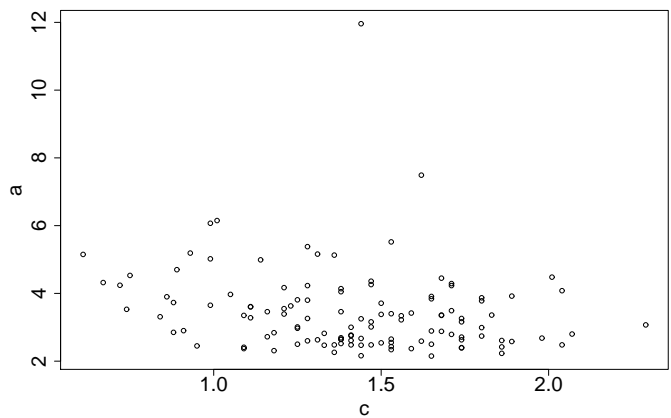}}
  \caption{Scatter plot of power law coefficients $a$ as a function of cluster concentration.\label{Exponentsc}}
\end{figure}

Figure \ref{HistExponentsPowerLaw} shows the histogram of the best-fit coefficient $a$ for the relation defined by Eq.~\ref{Eq:powerlawreal}, for the sample of $145$ GCs with sufficient-quality surface brightness profiles. The mean and median values of $a$ for the sample are $3.5$ and $3.2$ respectively. A value of $3$ would be expected for partially relaxed models produced by collisionless collapse (\cite{BeSt}, see also \cite{effenu}).
Figure \ref{Exponentsc} shows a scatter plot of $a$ versus the concentration parameter $c$; the values of $c$ are taken from \cite{CatMcLaughlin} and obtained by fitting King models to the surface brightness profiles. No significant trend emerges from the sample of $121$ GCs for which $a$ and $c$ are available.

\begin{figure}
  \resizebox{\hsize}{!}{\includegraphics[angle = 270]{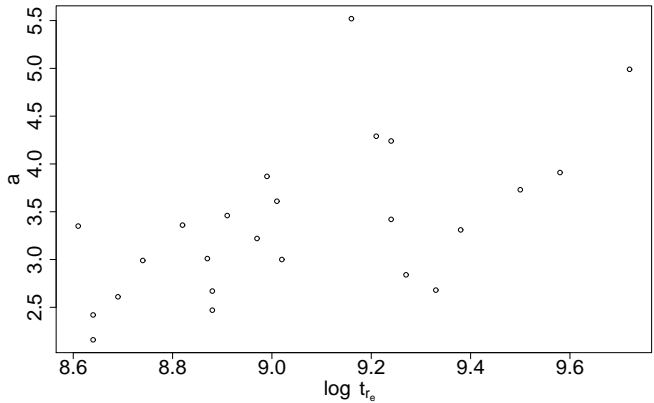}}
  \caption{Correlation of best-fit power law coefficients $a$ for the outer part of GC extended surface brightness profiles with half-light relaxation time by \cite{CatMcLaughlin}.\label{LogRelTimeVSalpha}}
\end{figure}

With the full sample of $145$ GCs, we do not identify any trend in the slopes represented by $a$ with either the half-light relaxation time or the ratio of GC age to relaxation time. This is unsurprising, since the surface brightness profiles of most clusters in this sample are not sufficiently extended in terms of the truncation radius (as obtained by fitting \cite{KingModels} models to the profile).
We then restrict our attention to a sample of $24$ GCs with a measured half-light relaxation time and an extended observed surface brightness profile \citep[][]{CatMcLaughlin}.
Figure \ref{LogRelTimeVSalpha} shows a trend in power-law exponent $a$ with log half-light relaxation time, which is quantitatively confirmed by a correlation coefficient $r = 0.58$ that is significant to $2 \times 10^{-3}$. The trend is mild but points to the existence of a relation between two-body relaxation phenomena and the shape of the outer envelope of GCs.

\section{Discussion and conclusions}
\label{Sect:DiscussionAndConclusions}
By considering a sample of $129$ Galactic and extragalactic GCs, we have found a linear relation between ${\mathit{SB}}_e$ and $\log R_e$, given by Eq.~(\ref{Eq:SBelogReRelat}). The relevant coefficient $5.25 \pm 0.44$ is compatible with $5$, i.e., with cluster absolute magnitude not showing a systematic linear trend with $\log R_e$. Indeed, when examined in the ($\log R_e$, $M_V$) plane, the cluster absolute magnitude shows no trend with size \citep[in contrast to the trend exhibited by dwarf spheroidal galaxies; see][]{vandenBerghNearbyDSPH}.

The typical scatter about the relation is $1$ mag in ${\mathit{SB}}_e$. Young clusters, of age below $4$ Gyr, exhibit a larger scatter of about $1.4$ mag. GC age seems to be an important factor in driving this scatter, because young clusters are found to display a clear trend of residuals to the relation with age, with a correlation coefficient of $r = 0.79$, while old GCs do not exhibit signs of such a correlation.
If we assume that all the massive young clusters included in our sample are genuine GC progenitors, our finding suggests that the scaling law being considered shows signs of evolution with age, i.e., that GCs of different ages differ with respect to this scaling law. Moreover, the larger scatter displayed by young clusters points to an evolutionary origin of the relation between ${\mathit{SB}}_e$ and $\log R_e$, as opposed to a primordial one.
This result is in contrast to the suggestion of \cite{Bellazzini} that ``at earlier times, globular clusters populated a line in the three-dimensional S-space, i.e., their original dynamical structure was fully determined by a single physical parameter''.

In principle, modeling the processes that generate the relation between ${\mathit{SB}}_e$ and $\log R_e$ would require taking into account passive stellar evolution, internal dynamical phenomena (evaporation and core-collapse driven by two-body relaxation), and dynamical interaction with the environment.
We find that the trend with age becomes stronger for young clusters if age is measured in units of the half-light relaxation time of each GC, i.e., if \emph{dynamical age} with respect to two-body relaxation is considered, the correlation coefficient rises to $0.90$. Given this result, we argue that two-body relaxation is the key ingredient in shaping the evolution of the ${\mathit{SB}}_e$-$\log R_e$ scaling law, at least for young clusters. This may not be the case for older clusters, which do not show such a trend. The evolution of these clusters in the ($\log R_e$, ${\mathit{SB}}_e$) plane is likely to be influenced by their host environment, but the available data is of limited use in resolving this issue. The distance to the host galaxy is only a simple proxy for the degree of dynamical disturbance that the GC suffers from it, while more detailed orbital data is available only for a sample of $48$ old, Galactic GCs \citep[][]{AMPOrbits48GC}. In any case, we find no trend between the residuals to the ${\mathit{SB}}_e$-$\log R_e$ relation and either distance from the host galaxy or pericenter distance and orbital eccentricity, when these data are available. When GC age is measured in units of the bulge-interaction and disk-interaction destruction times based on the orbital data provided by \cite{AMPOrbits48GC}, weak ($r \approx 0.5$) trends are found with ${\mathit{SB}}_e$-$\log R_e$ relation residuals. While this suggests that dynamical interaction with the galactic environment may exert an influence on GCs with observable effects in the ($\log R_e$, ${\mathit{SB}}_e$) plane, to draw conclusive evidence, the correlation with age alone should be removed from the trend and a larger sample of clusters with accurately measured orbits is necessary.

In contrast to what we noted for fundamental plane residuals in Paper I, we find here that central surface brightness profile slopes measured by \cite{NoyolaGalacticSlopes} and \cite{NoyolaExtragalacticSlopes} do not correlate with residuals from the ${\mathit{SB}}_e$-$\log R_e$ relation. This result suggests that velocity dispersion information is necessary to understand the mechanism producing cuspy GC surface brightness profiles.

\emph{Acknowledgments.} We wish to thank S. Degl'Innocenti, W. Harris, M. Lombardi, M. Trenti, E. Vesperini, and D. Kruijssen for their comments and helpful suggestions. This research has made use of the SIMBAD data-base, operated at CDS, Strasbourg, France. Part of this work was carried out at the Kavli Institute for Theoretical Physics, Santa Barbara (CA), while we participated in the Program \emph{Formation and Evolution of Globular Clusters}.

\bibliographystyle{aa}
\bibliography{fl}

\longtab{1}{
\begin{longtable}{llllll}
\caption{Adopted photometric quantities.\label{MeasuredPhotQTable}}\\
\hline\hline
ID & $m$ & $E(B - V)$ & ${|M - m|}_0$ & $\log {r_e}$ & $a$ \\
\hline
\endfirsthead
\caption{continued.}\\
\hline\hline
ID & $m$ & $E(B - V)$ & ${|M - m|}_0$ & $\log {r_e}$ & $a$ \\
\hline
\endhead
\hline\hline
\endfoot
ARP 2 & $12.74 \pm 0.03$ & $0.11 \pm 0.011$ & $17.37 \pm 0.18$ & $2.1 \pm 0.02$ & $2.85$ \\
FORNAX 1 & $15.44 \pm 0.08$ & $0.071 \pm 0.007$ & $20.68 \pm 0.22$ & $1.24 \pm 0.06$ & $5.15$ \\
FORNAX 2 & $13.64 \pm 0.06$ & $0.054 \pm 0.005$ & $20.68 \pm 0.22$ & $1.07 \pm 0.05$ & $3.31$ \\
FORNAX 3 & $12.73 \pm 0.09$ & $0.035 \pm 0.004$ & $20.68 \pm 0.21$ & $0.85 \pm 0.11$ & $7.49$ \\
FORNAX 4 & $13.56 \pm 0.13$ & $0.14 \pm 0.014$ & $20.68 \pm 0.24$ & $0.78 \pm 0.15$ & $3.01$ \\
FORNAX 5 & $13.52 \pm 0.08$ & $0.031 \pm 0.003$ & $20.68 \pm 0.21$ & $0.83 \pm 0.09$ & $3.78$ \\
IC 4499 & $9.91 \pm 0.02$ & $0.175 \pm 0.013$ & $16.53 \pm 0.1$ & $2.04 \pm 0.02$ & $4.17$ \\
LMC-HODGE 11 & $11.24 \pm 0.06$ & $0.046 \pm 0.005$ & $18.5 \pm 0.21$ & $1.49 \pm 0.05$ & $3.81$ \\
LMC-HODGE 14 & $12.88 \pm 0.14$ & $0.105 \pm 0.01$ & $18.5 \pm 0.23$ & $1.38 \pm 0.16$ & $3.34$ \\
LMC-NGC 1466 & $11.18 \pm 0.04$ & $0.09 \pm 0.009$ & $18.43 \pm 0.1$ & $1.3 \pm 0.03$ & $3.28$ \\
LMC-NGC 1711 & $9.57 \pm 0.06$ & $0.002 \pm 0.001$ & $18.5 \pm 0.2$ & $1.36 \pm 0.07$ & $3.25$ \\
LMC-NGC 1754 & $11.47 \pm 0.07$ & $0.096 \pm 0.01$ & $18.5 \pm 0.23$ & $1.19 \pm 0.09$ & $2.71$ \\
LMC-NGC 1777 & $11.44 \pm 0.06$ & $0 \pm 0.001$ & $18.5 \pm 0.2$ & $1.53 \pm 0.05$ & $3.26$ \\
LMC-NGC 1786 & $10.48 \pm 0.06$ & $0.125 \pm 0.012$ & $18.5 \pm 0.24$ & $1.26 \pm 0.07$ & $2.15$ \\
LMC-NGC 1805 & $10.29 \pm 0.07$ & $0.144 \pm 0.014$ & $18.5 \pm 0.24$ & $1.2 \pm 0.08$ & $2.6$ \\
LMC-NGC 1818 & $9.25 \pm 0.07$ & $0.15 \pm 0.015$ & $18.5 \pm 0.25$ & $1.38 \pm 0.08$ & $2.77$ \\
LMC-NGC 1831 & $10.4 \pm 0.05$ & $0.104 \pm 0.01$ & $18.5 \pm 0.23$ & $1.52 \pm 0.04$ & $2.84$ \\
LMC-NGC 1835 & $9.94 \pm 0.04$ & $0.105 \pm 0.01$ & $18.5 \pm 0.23$ & $1.02 \pm 0.04$ & $2.47$ \\
LMC-NGC 1847 & $10.01 \pm 0.13$ & $0.165 \pm 0.016$ & $18.5 \pm 0.25$ & $1.73 \pm 0.13$ & $3.07$ \\
LMC-NGC 1850 & $8.42 \pm 0.1$ & $0.102 \pm 0.01$ & $18.5 \pm 0.23$ & $1.59 \pm 0.11$ & $4.23$ \\
LMC-NGC 1856 & $9.28 \pm 0.1$ & $0.228 \pm 0.023$ & $18.5 \pm 0.27$ & $1.5 \pm 0.11$ & $3.16$ \\
LMC-NGC 1868 & $11.23 \pm 0.05$ & $0.132 \pm 0.013$ & $18.5 \pm 0.24$ & $1.18 \pm 0.05$ & $2.48$ \\
LMC-NGC 1898 & $10.91 \pm 0.07$ & $0.083 \pm 0.008$ & $18.5 \pm 0.23$ & $1.44 \pm 0.09$ & $4.36$ \\
LMC-NGC 1916 & $10.17 \pm 0.04$ & $0.188 \pm 0.019$ & $18.5 \pm 0.26$ & $0.97 \pm 0.04$ & $4.26$ \\
LMC-NGC 2004 & $9.15 \pm 0.09$ & $0.117 \pm 0.012$ & $18.5 \pm 0.24$ & $1.33 \pm 0.12$ & $3.84$ \\
LMC-NGC 2005 & $10.92 \pm 0.08$ & $0.12 \pm 0.012$ & $18.5 \pm 0.24$ & $1.07 \pm 0.09$ & $2.34$ \\
LMC-NGC 2011 & $9.76 \pm 0.14$ & $0.045 \pm 0.004$ & $18.5 \pm 0.21$ & $1.42 \pm 0.14$ & $4.08$ \\
LMC-NGC 2019 & $10.59 \pm 0.07$ & $0.138 \pm 0.014$ & $18.5 \pm 0.24$ & $1.09 \pm 0.1$ & $3.36$ \\
LMC-NGC 2031 & $10.05 \pm 0.08$ & $0.119 \pm 0.012$ & $18.5 \pm 0.24$ & $1.6 \pm 0.09$ & $3.38$ \\
LMC-NGC 2100 & $9.2 \pm 0.08$ & $0.208 \pm 0.021$ & $18.5 \pm 0.26$ & $1.31 \pm 0.11$ & $3.26$ \\
LMC-NGC 2121 & $11.02 \pm 0.12$ & $0.122 \pm 0.012$ & $18.5 \pm 0.24$ & $1.67 \pm 0.08$ & $4.32$ \\
LMC-NGC 2136 & $10.34 \pm 0.08$ & $0.189 \pm 0.019$ & $18.5 \pm 0.26$ & $1.17 \pm 0.07$ & $2.5$ \\
LMC-NGC 2153 & $13.11 \pm 0.3$ & $0.028 \pm 0.003$ & $18.5 \pm 0.21$ & $1.01 \pm 0.32$ & $11.96$ \\
LMC-NGC 2155 & $11.68 \pm 0.1$ & $0.079 \pm 0.008$ & $18.5 \pm 0.22$ & $1.63 \pm 0.09$ & $2.45$ \\
LMC-NGC 2157 & $9.56 \pm 0.06$ & $0.098 \pm 0.01$ & $18.5 \pm 0.23$ & $1.42 \pm 0.08$ & $2.67$ \\
LMC-NGC 2159 & $10.74 \pm 0.13$ & $0.188 \pm 0.019$ & $18.5 \pm 0.26$ & $1.5 \pm 0.19$ & $2.16$ \\
LMC-NGC 2162 & $12.32 \pm 0.11$ & $0.012 \pm 0.001$ & $18.5 \pm 0.2$ & $1.31 \pm 0.1$ & $5.38$ \\
LMC-NGC 2172 & $11.08 \pm 0.12$ & $0.092 \pm 0.009$ & $18.5 \pm 0.23$ & $1.52 \pm 0.15$ & $2.26$ \\
LMC-NGC 2173 & $11.53 \pm 0.09$ & $0.108 \pm 0.011$ & $18.5 \pm 0.23$ & $1.62 \pm 0.09$ & $2.65$ \\
LMC-NGC 2193 & $12.53 \pm 0.23$ & $0.044 \pm 0.004$ & $18.5 \pm 0.21$ & $1.39 \pm 0.24$ & $2.48$ \\
LMC-NGC 2210 & $10.75 \pm 0.05$ & $0.096 \pm 0.01$ & $18.5 \pm 0.23$ & $1.17 \pm 0.04$ & $3.61$ \\
LMC-NGC 2213 & $11.92 \pm 0.12$ & $0.021 \pm 0.002$ & $18.5 \pm 0.21$ & $1.34 \pm 0.13$ & $2.48$ \\
LMC-NGC 2214 & $10.05 \pm 0.09$ & $0.018 \pm 0.002$ & $18.5 \pm 0.21$ & $1.6 \pm 0.13$ & $2.63$ \\
LMC-NGC 2231 & $12.12 \pm 0.15$ & $0.106 \pm 0.011$ & $18.5 \pm 0.23$ & $1.48 \pm 0.12$ & $5.13$ \\
LMC-NGC 2249 & $11.9 \pm 0.15$ & $0.052 \pm 0.005$ & $18.5 \pm 0.22$ & $1.21 \pm 0.13$ & $3.55$ \\
LMC-NGC 2257 & $11.33 \pm 0.07$ & $0 \pm 0.001$ & $18.5 \pm 0.2$ & $1.7 \pm 0.04$ & $2.9$ \\
LMC-SL 842 & $13.42 \pm 0.22$ & $0.104 \pm 0.01$ & $18.5 \pm 0.23$ & $1.25 \pm 0.17$ & $6.15$ \\
NGC 104 & $4 \pm 0.01$ & $0.04 \pm 0.003$ & $13.29 \pm 0.06$ & $2.23 \pm 0.02$ & $4.48$ \\
NGC 1261 & $8.62 \pm 0.02$ & $0.012 \pm 0.001$ & $16.01 \pm 0.05$ & $1.62 \pm 0.02$ & $3.46$ \\
NGC 1851 & $7.24 \pm 0.04$ & $0.02 \pm 0.001$ & $15.42 \pm 0.05$ & $1.51 \pm 0.05$ & $2.42$ \\
NGC 1904 & $8.08 \pm 0.02$ & $0.01 \pm 0.001$ & $15.64 \pm 0.05$ & $1.62 \pm 0.02$ & $2.88$ \\
NGC 2808 & $6.38 \pm 0.01$ & $0.164 \pm 0.012$ & $15.09 \pm 0.1$ & $1.65 \pm 0.01$ & $3.22$ \\
NGC 288 & $8.21 \pm 0.02$ & $0.03 \pm 0.003$ & $14.67 \pm 0.11$ & $2.13 \pm 0.01$ & $6.07$ \\
NGC 3201 & $7.07 \pm 0.02$ & $0.227 \pm 0.016$ & $13.5 \pm 0.1$ & $2.18 \pm 0.01$ & $3.8$ \\
NGC 362 & $6.59 \pm 0.02$ & $0.004 \pm 0.001$ & $14.72 \pm 0.07$ & $1.62 \pm 0.03$ & $2.99$ \\
NGC 4147 & $10.29 \pm 0.02$ & $0.012 \pm 0.001$ & $16.4 \pm 0.07$ & $1.45 \pm 0.02$ & $3.36$ \\
NGC 4590 & $8.29 \pm 0.01$ & $0.044 \pm 0.003$ & $15.12 \pm 0.05$ & $1.93 \pm 0.01$ & $3$ \\
NGC 5024 & $7.69 \pm 0.01$ & $0.01 \pm 0.001$ & $16.35 \pm 0.08$ & $1.84 \pm 0.01$ & $3.91$ \\
NGC 5053 & $9.66 \pm 0.04$ & $0.03 \pm 0.003$ & $16.17 \pm 0.08$ & $2.21 \pm 0.02$ & $4.53$ \\
NGC 5272 & $6.53 \pm 0.02$ & $0.01 \pm 0.001$ & $14.99 \pm 0.05$ & $1.85 \pm 0.02$ & $3.92$ \\
NGC 5286 & $7.2 \pm 0.01$ & $0.24 \pm 0.024$ & $15.25 \pm 0.17$ & $1.72 \pm 0.02$ & $2.75$ \\
NGC 5466 & $9.48 \pm 0.02$ & $0 \pm 0.001$ & $16.12 \pm 0.1$ & $2.12 \pm 0.01$ & $3.73$ \\
NGC 5634 & $9.93 \pm 0.02$ & $0.01 \pm 0.001$ & $17.17 \pm 0.09$ & $1.61 \pm 0.02$ & $2.8$ \\
NGC 5694 & $10.12 \pm 0.01$ & $0.101 \pm 0.007$ & $17.77 \pm 0.09$ & $1.38 \pm 0.02$ & $2.58$ \\
NGC 5824 & $8.84 \pm 0.01$ & $0.128 \pm 0.009$ & $17.6 \pm 0.08$ & $1.44 \pm 0.01$ & $2.68$ \\
NGC 5897 & $8.67 \pm 0.02$ & $0.08 \pm 0.008$ & $15.61 \pm 0.09$ & $2.11 \pm 0.01$ & $3.9$ \\
NGC 5904 & $5.86 \pm 0.01$ & $0.03 \pm 0.002$ & $14.37 \pm 0.05$ & $2.04 \pm 0.01$ & $4.29$ \\
NGC 5986 & $7.81 \pm 0.01$ & $0.27 \pm 0.027$ & $15.39 \pm 0.16$ & $1.73 \pm 0.01$ & $3.63$ \\
NGC 6093 & $7.45 \pm 0.01$ & $0.196 \pm 0.014$ & $15.05 \pm 0.09$ & $1.59 \pm 0.01$ & $3.35$ \\
NGC 6121 & $5.67 \pm 0.01$ & $0.36 \pm 0.036$ & $11.62 \pm 0.21$ & $2.48 \pm 0.01$ & $2.89$ \\
NGC 6139 & $8.86 \pm 0.02$ & $0.72 \pm 0.072$ & $15.45 \pm 0.3$ & $1.87 \pm 0.02$ & $2.23$ \\
NGC 6171 & $8.28 \pm 0.01$ & $0.38 \pm 0.027$ & $13.81 \pm 0.15$ & $2.1 \pm 0.01$ & $2.43$ \\
NGC 6205 & $5.89 \pm 0.01$ & $0.012 \pm 0.001$ & $14.46 \pm 0.07$ & $1.99 \pm 0.01$ & $5.52$ \\
NGC 6218 & $7.17 \pm 0.01$ & $0.189 \pm 0.013$ & $13.58 \pm 0.16$ & $2.04 \pm 0.01$ & $4.23$ \\
NGC 6229 & $9.56 \pm 0.01$ & $0.012 \pm 0.001$ & $17.42 \pm 0.05$ & $1.35 \pm 0.01$ & $3.71$ \\
NGC 6254 & $6.66 \pm 0.01$ & $0.28 \pm 0.028$ & $13.35 \pm 0.19$ & $2.04 \pm 0.01$ & $4.05$ \\
NGC 6256 & $11.24 \pm 0.02$ & $1.18 \pm 0.118$ & $14 \pm 0.43$ & $1.68 \pm 0.01$ & $3.41$ \\
NGC 6266 & $6.46 \pm 0.02$ & $0.5 \pm 0.035$ & $14.15 \pm 0.2$ & $1.81 \pm 0.03$ & $2.79$ \\
NGC 6273 & $6.74 \pm 0.01$ & $0.34 \pm 0.034$ & $15.05 \pm 0.18$ & $1.96 \pm 0.01$ & $2.55$ \\
NGC 6284 & $8.7 \pm 0.02$ & $0.29 \pm 0.029$ & $16.06 \pm 0.16$ & $1.7 \pm 0.01$ & $2.48$ \\
NGC 6287 & $9.46 \pm 0.01$ & $0.66 \pm 0.066$ & $14.79 \pm 0.27$ & $1.71 \pm 0.01$ & $2.64$ \\
NGC 6293 & $7.97 \pm 0.02$ & $0.62 \pm 0.062$ & $14.22 \pm 0.26$ & $1.92 \pm 0.02$ & $2.16$ \\
NGC 6316 & $9.02 \pm 0.02$ & $0.51 \pm 0.051$ & $15.59 \pm 0.25$ & $1.74 \pm 0.02$ & $2.5$ \\
NGC 6325 & $10.49 \pm 0.01$ & $1.03 \pm 0.103$ & $14.42 \pm 0.39$ & $1.74 \pm 0.01$ & $2.3$ \\
NGC 6333 & $7.75 \pm 0.01$ & $0.36 \pm 0.036$ & $14.62 \pm 0.26$ & $1.77 \pm 0.01$ & $3.01$ \\
NGC 6341 & $6.56 \pm 0.01$ & $0.02 \pm 0.002$ & $14.74 \pm 0.11$ & $1.75 \pm 0.01$ & $4.45$ \\
NGC 6342 & $9.78 \pm 0.02$ & $0.46 \pm 0.046$ & $14.95 \pm 0.22$ & $1.65 \pm 0.02$ & $3.15$ \\
NGC 6356 & $8.13 \pm 0.01$ & $0.28 \pm 0.028$ & $16 \pm 0.17$ & $1.79 \pm 0.01$ & $2.37$ \\
NGC 6362 & $7.54 \pm 0.01$ & $0.08 \pm 0.008$ & $14.48 \pm 0.11$ & $2.16 \pm 0.01$ & $2.37$ \\
NGC 6366 & $8.68 \pm 0.01$ & $0.69 \pm 0.069$ & $12.88 \pm 0.31$ & $2.28 \pm 0.01$ & $3.53$ \\
NGC 6388 & $6.87 \pm 0.01$ & $0.37 \pm 0.037$ & $15.42 \pm 0.2$ & $1.51 \pm 0.01$ & $3.49$ \\
NGC 6397 & $5.95 \pm 0.01$ & $0.18 \pm 0.018$ & $11.89 \pm 0.16$ & $2.24 \pm 0.02$ & $3.14$ \\
NGC 6401 & $9.57 \pm 0.02$ & $0.91 \pm 0.091$ & $14.55 \pm 0.41$ & $1.84 \pm 0.02$ & $2.67$ \\
NGC 6402 & $7.73 \pm 0.01$ & $0.66 \pm 0.066$ & $14.91 \pm 0.28$ & $1.91 \pm 0.01$ & $3.65$ \\
NGC 6440 & $9.43 \pm 0.01$ & $1.09 \pm 0.077$ & $14.69 \pm 0.32$ & $1.51 \pm 0.01$ & $2.59$ \\
NGC 6441 & $7.19 \pm 0.03$ & $0.47 \pm 0.047$ & $15.77 \pm 0.23$ & $1.61 \pm 0.04$ & $2.4$ \\
NGC 6517 & $10.54 \pm 0.01$ & $1.23 \pm 0.123$ & $15.04 \pm 0.45$ & $1.58 \pm 0.01$ & $2.56$ \\
NGC 6522 & $8.46 \pm 0.03$ & $0.54 \pm 0.054$ & $14.66 \pm 0.25$ & $1.63 \pm 0.03$ & $8.3$ \\
NGC 6528 & $9.59 \pm 0.06$ & $0.62 \pm 0.062$ & $14.35 \pm 0.39$ & $1.43 \pm 0.05$ & $2.54$ \\
NGC 6539 & $9.64 \pm 0.03$ & $0.97 \pm 0.097$ & $14.9 \pm 0.37$ & $2.06 \pm 0.04$ & $2.39$ \\
NGC 6553 & $8.03 \pm 0.01$ & $0.84 \pm 0.084$ & $13.44 \pm 0.46$ & $1.88 \pm 0.01$ & $2.72$ \\
NGC 6569 & $8.8 \pm 0.02$ & $0.57 \pm 0.057$ & $15.26 \pm 0.25$ & $1.76 \pm 0.02$ & $2.63$ \\
NGC 6584 & $8.93 \pm 0.01$ & $0.023 \pm 0.002$ & $15.76 \pm 0.06$ & $1.71 \pm 0.01$ & $3.16$ \\
NGC 6624 & $7.82 \pm 0.01$ & $0.28 \pm 0.028$ & $14.43 \pm 0.18$ & $1.79 \pm 0.01$ & $2.25$ \\
NGC 6637 & $7.46 \pm 0.01$ & $0.17 \pm 0.012$ & $14.7 \pm 0.1$ & $1.8 \pm 0.01$ & $2.52$ \\
NGC 6638 & $8.89 \pm 0.01$ & $0.39 \pm 0.039$ & $15.14 \pm 0.21$ & $1.53 \pm 0.01$ & $2.82$ \\
NGC 6642 & $9.35 \pm 0.03$ & $0.44 \pm 0.044$ & $14.83 \pm 0.22$ & $1.63 \pm 0.02$ & $2.59$ \\
NGC 6652 & $8.98 \pm 0.02$ & $0.103 \pm 0.007$ & $15.01 \pm 0.09$ & $1.53 \pm 0.01$ & $2.74$ \\
NGC 6681 & $7.93 \pm 0.02$ & $0.064 \pm 0.005$ & $15 \pm 0.08$ & $1.78 \pm 0.02$ & $2.44$ \\
NGC 6712 & $8.11 \pm 0.03$ & $0.419 \pm 0.03$ & $14.27 \pm 0.15$ & $1.93 \pm 0.03$ & $3.97$ \\
NGC 6723 & $7.24 \pm 0.02$ & $0.05 \pm 0.005$ & $14.74 \pm 0.11$ & $1.95 \pm 0.02$ & $3.6$ \\
NGC 6752 & $5.67 \pm 0.01$ & $0.04 \pm 0.004$ & $13.13 \pm 0.16$ & $2.08 \pm 0.02$ & $3.84$ \\
NGC 6809 & $6.84 \pm 0.01$ & $0.07 \pm 0.007$ & $13.78 \pm 0.12$ & $2.22 \pm 0.02$ & $5.19$ \\
NGC 6864 & $8.52 \pm 0.02$ & $0.21 \pm 0.021$ & $16.49 \pm 0.15$ & $1.44 \pm 0.02$ & $3.87$ \\
NGC 6934 & $8.77 \pm 0.01$ & $0.09 \pm 0.006$ & $16.05 \pm 0.08$ & $1.56 \pm 0.01$ & $3.4$ \\
NGC 6981 & $9.3 \pm 0.02$ & $0.05 \pm 0.004$ & $16.1 \pm 0.06$ & $1.71 \pm 0.02$ & $3.39$ \\
NGC 7006 & $10.63 \pm 0.01$ & $0.05 \pm 0.005$ & $18.03 \pm 0.17$ & $1.41 \pm 0.01$ & $2.61$ \\
NGC 7078 & $6.38 \pm 0.01$ & $0.09 \pm 0.006$ & $15.15 \pm 0.08$ & $1.8 \pm 0.01$ & $3.18$ \\
NGC 7089 & $6.41 \pm 0.01$ & $0.01 \pm 0.001$ & $15.47 \pm 0.09$ & $1.78 \pm 0.01$ & $3.42$ \\
NGC 7099 & $7.57 \pm 0.01$ & $0.03 \pm 0.002$ & $14.7 \pm 0.07$ & $1.8 \pm 0.01$ & $2.92$ \\
NGC 7492 & $11.39 \pm 0.03$ & $0 \pm 0.001$ & $17.1 \pm 0.1$ & $1.84 \pm 0.02$ & $4.24$ \\
SMC-KRON 3 & $11.1 \pm 0.07$ & $0.016 \pm 0.002$ & $18.89 \pm 0.2$ & $1.59 \pm 0.06$ & $4.99$ \\
SMC-NGC 121 & $10.71 \pm 0.05$ & $0.146 \pm 0.015$ & $18.89 \pm 0.25$ & $1.37 \pm 0.06$ & $2.31$ \\
SMC-NGC 152 & $11.65 \pm 0.08$ & $0.12 \pm 0.012$ & $18.89 \pm 0.24$ & $1.7 \pm 0.06$ & $3.35$ \\
SMC-NGC 176 & $12.21 \pm 0.19$ & $0 \pm 0.001$ & $18.89 \pm 0.2$ & $1.34 \pm 0.18$ & $2.97$ \\
SMC-NGC 361 & $11.13 \pm 0.09$ & $0.069 \pm 0.007$ & $18.89 \pm 0.22$ & $1.55 \pm 0.06$ & $5.02$ \\
SMC-NGC 411 & $11.61 \pm 0.11$ & $0.054 \pm 0.005$ & $18.89 \pm 0.22$ & $1.42 \pm 0.1$ & $2.69$ \\
SMC-NGC 416 & $11.18 \pm 0.06$ & $0.127 \pm 0.013$ & $18.89 \pm 0.24$ & $1.24 \pm 0.05$ & $4.7$ \\
SMC-NGC 458 & $11.09 \pm 0.09$ & $0.022 \pm 0.002$ & $18.89 \pm 0.21$ & $1.44 \pm 0.07$ & $2.41$ \\
\end{longtable}
 \noindent $^a$ cluster name\\ 
 $^b$ V-band integrated apparent magnitude, this paper, based on model-independent smoothing and integration of \cite{CatTrager} surface brightness profiles, mag\\
 $^c$ line-of-sight reddening, \cite{FerraroDistances} and \cite{RecioBlancoDistances} (galactic GCs), \cite{RecioBlancoDistances} (extragalactic clusters), mag\\
$^d$ true distance modulus \cite{FerraroDistances} and \cite{RecioBlancoDistances} (galactic GCs), \cite{RecioBlancoDistances} (extragalactic clusters), mag\\
 $^e$ log projected angular half-light radius, this paper, based on model-independent smoothing and integration of \cite{CatTrager} surface brightness profiles, arcsec\\
 $^f$ slope of the outermost surface brightness profile, this paper, dimensionless\\
}

\longtab{2}{
\begin{longtable}{lllll}
\caption{Half-light photometric quantities ($a$ and $b$), integrated absolute magnitudes ($c$) and residuals ($d$) to the ${\mathit{SB}}_e$-$\log R_e$ relation.\label{DerivedPhotQTable}}\\
\hline\hline
ID & ${\mathit{SB}}_e$ & $\log R_e$ & $M$ & residuals\\
\hline
\endfirsthead
\caption{continued.}\\
\hline\hline
ID & ${\mathit{SB}}_e$ & $\log R_e$ & $M$ & residuals\\
\hline
\endhead
\hline\hline
\endfoot
ARP 2 & $24.87 \pm 0.16$ & $1.26 \pm 0.06$ & $-4.97 \pm 0.25$ & $2.67$ \\
FORNAX 1 & $23.38 \pm 0.4$ & $1.06 \pm 0.1$ & $-5.46 \pm 0.33$ & $2.23$ \\
FORNAX 2 & $20.83 \pm 0.33$ & $0.9 \pm 0.09$ & $-7.21 \pm 0.29$ & $0.52$ \\
FORNAX 3 & $18.86 \pm 0.65$ & $0.68 \pm 0.15$ & $-8.06 \pm 0.31$ & $-0.27$ \\
FORNAX 4 & $19.03 \pm 0.93$ & $0.61 \pm 0.2$ & $-7.55 \pm 0.42$ & $0.25$ \\
FORNAX 5 & $19.55 \pm 0.53$ & $0.65 \pm 0.13$ & $-7.26 \pm 0.29$ & $0.54$ \\
IC 4499 & $21.55 \pm 0.16$ & $1.03 \pm 0.04$ & $-7.16 \pm 0.16$ & $0.53$ \\
LMC-HODGE 11 & $20.54 \pm 0.33$ & $0.88 \pm 0.09$ & $-7.4 \pm 0.29$ & $0.33$ \\
LMC-HODGE 14 & $21.44 \pm 0.97$ & $0.77 \pm 0.21$ & $-5.95 \pm 0.4$ & $1.82$ \\
LMC-NGC 1466 & $19.39 \pm 0.22$ & $0.68 \pm 0.05$ & $-7.53 \pm 0.17$ & $0.26$ \\
LMC-NGC 1711 & $18.35 \pm 0.41$ & $0.75 \pm 0.11$ & $-8.94 \pm 0.26$ & $-1.17$ \\
LMC-NGC 1754 & $19.13 \pm 0.55$ & $0.58 \pm 0.14$ & $-7.33 \pm 0.33$ & $0.48$ \\
LMC-NGC 1777 & $21.08 \pm 0.31$ & $0.92 \pm 0.09$ & $-7.06 \pm 0.26$ & $0.67$ \\
LMC-NGC 1786 & $18.4 \pm 0.45$ & $0.65 \pm 0.12$ & $-8.41 \pm 0.34$ & $-0.62$ \\
LMC-NGC 1805 & $17.83 \pm 0.52$ & $0.59 \pm 0.13$ & $-8.66 \pm 0.36$ & $-0.85$ \\
LMC-NGC 1818 & $17.67 \pm 0.52$ & $0.77 \pm 0.13$ & $-9.71 \pm 0.36$ & $-1.95$ \\
LMC-NGC 1831 & $19.67 \pm 0.28$ & $0.91 \pm 0.09$ & $-8.42 \pm 0.31$ & $-0.69$ \\
LMC-NGC 1835 & $16.69 \pm 0.27$ & $0.41 \pm 0.09$ & $-8.89 \pm 0.31$ & $-1.03$ \\
LMC-NGC 1847 & $20.13 \pm 0.83$ & $1.12 \pm 0.18$ & $-9 \pm 0.43$ & $-1.33$ \\
LMC-NGC 1850 & $18.04 \pm 0.68$ & $0.98 \pm 0.16$ & $-10.4 \pm 0.36$ & $-2.69$ \\
LMC-NGC 1856 & $18.05 \pm 0.72$ & $0.89 \pm 0.16$ & $-9.93 \pm 0.45$ & $-2.19$ \\
LMC-NGC 1868 & $18.71 \pm 0.34$ & $0.57 \pm 0.1$ & $-7.68 \pm 0.33$ & $0.13$ \\
LMC-NGC 1898 & $19.84 \pm 0.55$ & $0.83 \pm 0.14$ & $-7.85 \pm 0.32$ & $-0.1$ \\
LMC-NGC 1916 & $16.44 \pm 0.29$ & $0.36 \pm 0.09$ & $-8.91 \pm 0.35$ & $-1.05$ \\
LMC-NGC 2004 & $17.42 \pm 0.73$ & $0.72 \pm 0.17$ & $-9.71 \pm 0.36$ & $-1.94$ \\
LMC-NGC 2005 & $17.87 \pm 0.57$ & $0.46 \pm 0.14$ & $-7.95 \pm 0.35$ & $-0.11$ \\
LMC-NGC 2011 & $18.71 \pm 0.85$ & $0.81 \pm 0.18$ & $-8.88 \pm 0.37$ & $-1.13$ \\
LMC-NGC 2019 & $17.58 \pm 0.61$ & $0.48 \pm 0.15$ & $-8.34 \pm 0.36$ & $-0.5$ \\
LMC-NGC 2031 & $19.67 \pm 0.57$ & $0.99 \pm 0.14$ & $-8.82 \pm 0.36$ & $-1.11$ \\
LMC-NGC 2100 & $17.08 \pm 0.7$ & $0.7 \pm 0.16$ & $-9.95 \pm 0.41$ & $-2.16$ \\
LMC-NGC 2121 & $20.98 \pm 0.56$ & $1.06 \pm 0.13$ & $-7.86 \pm 0.4$ & $-0.17$ \\
LMC-NGC 2136 & $17.59 \pm 0.49$ & $0.56 \pm 0.12$ & $-8.75 \pm 0.4$ & $-0.93$ \\
LMC-NGC 2153 & $20.04 \pm 1.91$ & $0.4 \pm 0.36$ & $-5.48 \pm 0.52$ & $2.38$ \\
LMC-NGC 2155 & $21.58 \pm 0.58$ & $1.02 \pm 0.13$ & $-7.06 \pm 0.35$ & $0.64$ \\
LMC-NGC 2157 & $18.35 \pm 0.5$ & $0.81 \pm 0.13$ & $-9.24 \pm 0.33$ & $-1.49$ \\
LMC-NGC 2159 & $19.65 \pm 1.13$ & $0.89 \pm 0.24$ & $-8.34 \pm 0.44$ & $-0.61$ \\
LMC-NGC 2162 & $20.82 \pm 0.62$ & $0.7 \pm 0.14$ & $-6.22 \pm 0.32$ & $1.56$ \\
LMC-NGC 2172 & $20.39 \pm 0.9$ & $0.91 \pm 0.2$ & $-7.7 \pm 0.38$ & $0.02$ \\
LMC-NGC 2173 & $21.28 \pm 0.57$ & $1.01 \pm 0.14$ & $-7.31 \pm 0.35$ & $0.4$ \\
LMC-NGC 2193 & $21.33 \pm 1.44$ & $0.78 \pm 0.28$ & $-6.11 \pm 0.46$ & $1.65$ \\
LMC-NGC 2210 & $18.29 \pm 0.28$ & $0.56 \pm 0.09$ & $-8.05 \pm 0.31$ & $-0.23$ \\
LMC-NGC 2213 & $20.54 \pm 0.77$ & $0.73 \pm 0.17$ & $-6.65 \pm 0.33$ & $1.13$ \\
LMC-NGC 2214 & $19.98 \pm 0.75$ & $0.99 \pm 0.17$ & $-8.51 \pm 0.3$ & $-0.8$ \\
LMC-NGC 2231 & $21.19 \pm 0.78$ & $0.87 \pm 0.17$ & $-6.71 \pm 0.41$ & $1.03$ \\
LMC-NGC 2249 & $19.78 \pm 0.82$ & $0.6 \pm 0.17$ & $-6.76 \pm 0.38$ & $1.04$ \\
LMC-NGC 2257 & $21.83 \pm 0.27$ & $1.09 \pm 0.08$ & $-7.17 \pm 0.27$ & $0.51$ \\
LMC-SL 842 & $21.34 \pm 1.1$ & $0.64 \pm 0.22$ & $-5.4 \pm 0.48$ & $2.39$ \\
NGC 104 & $17.02 \pm 0.12$ & $0.58 \pm 0.03$ & $-9.42 \pm 0.08$ & $-1.6$ \\
NGC 1261 & $18.67 \pm 0.12$ & $0.51 \pm 0.03$ & $-7.43 \pm 0.07$ & $0.4$ \\
NGC 1851 & $16.72 \pm 0.29$ & $0.29 \pm 0.06$ & $-8.24 \pm 0.09$ & $-0.36$ \\
NGC 1904 & $18.12 \pm 0.13$ & $0.43 \pm 0.03$ & $-7.59 \pm 0.08$ & $0.26$ \\
NGC 2808 & $16.11 \pm 0.1$ & $0.36 \pm 0.03$ & $-9.22 \pm 0.14$ & $-1.35$ \\
NGC 288 & $20.78 \pm 0.08$ & $0.76 \pm 0.03$ & $-6.55 \pm 0.14$ & $1.21$ \\
NGC 3201 & $19.25 \pm 0.12$ & $0.57 \pm 0.03$ & $-7.13 \pm 0.17$ & $0.68$ \\
NGC 362 & $16.67 \pm 0.17$ & $0.25 \pm 0.04$ & $-8.15 \pm 0.09$ & $-0.25$ \\
NGC 4147 & $19.51 \pm 0.12$ & $0.42 \pm 0.03$ & $-6.15 \pm 0.08$ & $1.7$ \\
NGC 4590 & $19.81 \pm 0.07$ & $0.65 \pm 0.02$ & $-6.97 \pm 0.07$ & $0.82$ \\
NGC 5024 & $18.82 \pm 0.07$ & $0.8 \pm 0.03$ & $-8.69 \pm 0.1$ & $-0.94$ \\
NGC 5053 & $22.63 \pm 0.15$ & $1.14 \pm 0.04$ & $-6.6 \pm 0.13$ & $1.07$ \\
NGC 5272 & $17.73 \pm 0.12$ & $0.54 \pm 0.03$ & $-8.49 \pm 0.07$ & $-0.67$ \\
NGC 5286 & $17.05 \pm 0.19$ & $0.46 \pm 0.05$ & $-8.79 \pm 0.26$ & $-0.95$ \\
NGC 5466 & $22.06 \pm 0.07$ & $1.03 \pm 0.03$ & $-6.64 \pm 0.12$ & $1.06$ \\
NGC 5634 & $19.92 \pm 0.12$ & $0.73 \pm 0.04$ & $-7.27 \pm 0.11$ & $0.5$ \\
NGC 5694 & $18.69 \pm 0.14$ & $0.62 \pm 0.04$ & $-7.96 \pm 0.12$ & $-0.16$ \\
NGC 5824 & $17.65 \pm 0.09$ & $0.65 \pm 0.03$ & $-9.16 \pm 0.12$ & $-1.37$ \\
NGC 5897 & $20.96 \pm 0.1$ & $0.92 \pm 0.03$ & $-7.19 \pm 0.14$ & $0.54$ \\
NGC 5904 & $17.95 \pm 0.07$ & $0.6 \pm 0.02$ & $-8.6 \pm 0.07$ & $-0.8$ \\
NGC 5986 & $17.6 \pm 0.14$ & $0.5 \pm 0.04$ & $-8.41 \pm 0.25$ & $-0.58$ \\
NGC 6093 & $16.77 \pm 0.11$ & $0.29 \pm 0.03$ & $-8.2 \pm 0.15$ & $-0.32$ \\
NGC 6121 & $18.97 \pm 0.17$ & $0.5 \pm 0.05$ & $-7.07 \pm 0.33$ & $0.76$ \\
NGC 6139 & $17.97 \pm 0.34$ & $0.65 \pm 0.08$ & $-8.82 \pm 0.54$ & $-1.03$ \\
NGC 6171 & $19.61 \pm 0.15$ & $0.56 \pm 0.04$ & $-6.71 \pm 0.25$ & $1.11$ \\
NGC 6205 & $17.79 \pm 0.06$ & $0.57 \pm 0.02$ & $-8.61 \pm 0.09$ & $-0.79$ \\
NGC 6218 & $18.8 \pm 0.1$ & $0.45 \pm 0.04$ & $-7 \pm 0.21$ & $0.85$ \\
NGC 6229 & $18.25 \pm 0.07$ & $0.52 \pm 0.02$ & $-7.9 \pm 0.06$ & $-0.08$ \\
NGC 6254 & $18 \pm 0.15$ & $0.4 \pm 0.05$ & $-7.56 \pm 0.29$ & $0.3$ \\
NGC 6256 & $17.99 \pm 0.43$ & $0.17 \pm 0.1$ & $-6.42 \pm 0.82$ & $1.5$ \\
NGC 6266 & $15.95 \pm 0.28$ & $0.33 \pm 0.07$ & $-9.24 \pm 0.33$ & $-1.37$ \\
NGC 6273 & $17.47 \pm 0.17$ & $0.66 \pm 0.05$ & $-9.37 \pm 0.3$ & $-1.58$ \\
NGC 6284 & $18.28 \pm 0.16$ & $0.6 \pm 0.04$ & $-8.26 \pm 0.27$ & $-0.45$ \\
NGC 6287 & $17.94 \pm 0.27$ & $0.36 \pm 0.06$ & $-7.37 \pm 0.49$ & $0.49$ \\
NGC 6293 & $17.63 \pm 0.31$ & $0.45 \pm 0.07$ & $-8.17 \pm 0.47$ & $-0.33$ \\
NGC 6316 & $18.12 \pm 0.28$ & $0.55 \pm 0.07$ & $-8.15 \pm 0.43$ & $-0.33$ \\
NGC 6325 & $17.98 \pm 0.38$ & $0.31 \pm 0.09$ & $-7.13 \pm 0.72$ & $0.75$ \\
NGC 6333 & $17.46 \pm 0.17$ & $0.38 \pm 0.06$ & $-7.99 \pm 0.39$ & $-0.13$ \\
NGC 6341 & $17.23 \pm 0.06$ & $0.39 \pm 0.03$ & $-8.24 \pm 0.12$ & $-0.38$ \\
NGC 6342 & $18.58 \pm 0.26$ & $0.33 \pm 0.06$ & $-6.59 \pm 0.38$ & $1.28$ \\
NGC 6356 & $18.19 \pm 0.09$ & $0.68 \pm 0.03$ & $-8.74 \pm 0.26$ & $-0.95$ \\
NGC 6362 & $20.07 \pm 0.09$ & $0.74 \pm 0.03$ & $-7.19 \pm 0.15$ & $0.58$ \\
NGC 6366 & $19.94 \pm 0.27$ & $0.55 \pm 0.07$ & $-6.34 \pm 0.53$ & $1.48$ \\
NGC 6388 & $15.26 \pm 0.17$ & $0.28 \pm 0.05$ & $-9.69 \pm 0.32$ & $-1.81$ \\
NGC 6397 & $18.58 \pm 0.17$ & $0.31 \pm 0.05$ & $-6.5 \pm 0.23$ & $1.38$ \\
NGC 6401 & $17.93 \pm 0.4$ & $0.44 \pm 0.1$ & $-7.8 \pm 0.72$ & $0.04$ \\
NGC 6402 & $17.22 \pm 0.22$ & $0.58 \pm 0.06$ & $-9.22 \pm 0.5$ & $-1.41$ \\
NGC 6440 & $15.61 \pm 0.3$ & $0.14 \pm 0.07$ & $-8.64 \pm 0.57$ & $-0.72$ \\
NGC 6441 & $15.78 \pm 0.37$ & $0.46 \pm 0.09$ & $-10.03 \pm 0.4$ & $-2.19$ \\
NGC 6517 & $16.6 \pm 0.44$ & $0.28 \pm 0.1$ & $-8.32 \pm 0.84$ & $-0.43$ \\
NGC 6522 & $16.93 \pm 0.35$ & $0.25 \pm 0.08$ & $-7.88 \pm 0.44$ & $0.01$ \\
NGC 6528 & $16.8 \pm 0.5$ & $-0.01 \pm 0.13$ & $-6.68 \pm 0.64$ & $1.28$ \\
NGC 6539 & $18.92 \pm 0.53$ & $0.73 \pm 0.11$ & $-8.26 \pm 0.7$ & $-0.49$ \\
NGC 6553 & $16.81 \pm 0.32$ & $0.26 \pm 0.1$ & $-8.01 \pm 0.73$ & $-0.12$ \\
NGC 6569 & $17.81 \pm 0.3$ & $0.5 \pm 0.07$ & $-8.22 \pm 0.44$ & $-0.39$ \\
NGC 6584 & $19.4 \pm 0.07$ & $0.55 \pm 0.02$ & $-6.9 \pm 0.08$ & $0.92$ \\
NGC 6624 & $17.88 \pm 0.15$ & $0.36 \pm 0.05$ & $-7.48 \pm 0.28$ & $0.39$ \\
NGC 6637 & $17.92 \pm 0.1$ & $0.43 \pm 0.03$ & $-7.76 \pm 0.15$ & $0.08$ \\
NGC 6638 & $17.31 \pm 0.18$ & $0.24 \pm 0.05$ & $-7.46 \pm 0.34$ & $0.44$ \\
NGC 6642 & $18.12 \pm 0.26$ & $0.28 \pm 0.06$ & $-6.85 \pm 0.38$ & $1.04$ \\
NGC 6652 & $18.3 \pm 0.09$ & $0.22 \pm 0.03$ & $-6.35 \pm 0.12$ & $1.55$ \\
NGC 6681 & $18.64 \pm 0.13$ & $0.47 \pm 0.04$ & $-7.27 \pm 0.11$ & $0.57$ \\
NGC 6712 & $18.47 \pm 0.28$ & $0.48 \pm 0.06$ & $-7.46 \pm 0.27$ & $0.38$ \\
NGC 6723 & $18.82 \pm 0.14$ & $0.59 \pm 0.04$ & $-7.66 \pm 0.16$ & $0.15$ \\
NGC 6752 & $17.94 \pm 0.12$ & $0.4 \pm 0.05$ & $-7.58 \pm 0.19$ & $0.27$ \\
NGC 6809 & $19.72 \pm 0.14$ & $0.67 \pm 0.04$ & $-7.16 \pm 0.16$ & $0.63$ \\
NGC 6864 & $17.06 \pm 0.18$ & $0.43 \pm 0.05$ & $-8.62 \pm 0.23$ & $-0.77$ \\
NGC 6934 & $18.27 \pm 0.08$ & $0.46 \pm 0.03$ & $-7.56 \pm 0.11$ & $0.28$ \\
NGC 6981 & $19.68 \pm 0.13$ & $0.62 \pm 0.03$ & $-6.95 \pm 0.09$ & $0.85$ \\
NGC 7006 & $19.51 \pm 0.08$ & $0.71 \pm 0.04$ & $-7.56 \pm 0.19$ & $0.22$ \\
NGC 7078 & $17.08 \pm 0.08$ & $0.52 \pm 0.03$ & $-9.05 \pm 0.11$ & $-1.23$ \\
NGC 7089 & $17.26 \pm 0.06$ & $0.56 \pm 0.03$ & $-9.09 \pm 0.1$ & $-1.28$ \\
NGC 7099 & $18.47 \pm 0.07$ & $0.43 \pm 0.02$ & $-7.22 \pm 0.09$ & $0.62$ \\
NGC 7492 & $22.57 \pm 0.13$ & $0.95 \pm 0.04$ & $-5.71 \pm 0.13$ & $2.01$ \\
SMC-KRON 3 & $21 \pm 0.37$ & $1.06 \pm 0.1$ & $-7.84 \pm 0.28$ & $-0.15$ \\
SMC-NGC 121 & $19.1 \pm 0.4$ & $0.84 \pm 0.11$ & $-8.63 \pm 0.34$ & $-0.89$ \\
SMC-NGC 152 & $21.77 \pm 0.42$ & $1.17 \pm 0.11$ & $-7.61 \pm 0.36$ & $0.05$ \\
SMC-NGC 176 & $20.9 \pm 1.09$ & $0.81 \pm 0.22$ & $-6.68 \pm 0.39$ & $1.07$ \\
SMC-NGC 361 & $20.67 \pm 0.41$ & $1.02 \pm 0.1$ & $-7.97 \pm 0.33$ & $-0.27$ \\
SMC-NGC 411 & $20.53 \pm 0.62$ & $0.89 \pm 0.14$ & $-7.45 \pm 0.34$ & $0.29$ \\
SMC-NGC 416 & $18.98 \pm 0.35$ & $0.71 \pm 0.1$ & $-8.1 \pm 0.34$ & $-0.33$ \\
SMC-NGC 458 & $20.21 \pm 0.45$ & $0.91 \pm 0.11$ & $-7.87 \pm 0.3$ & $-0.14$ \\
\end{longtable}
\noindent$^a$ average V-band surface brightness within the half-light radius, this paper, mag/arcsec$^2$\\
$^b$ projected half-light radius, this paper, parsec\\
$^c$ V-band integrated absolute magnitude, this paper, mag\\
$^d$ residuals in ${\mathit{SB}}_e$ to the ${\mathit{SB}}_e$-$\log R_e$ relation, this paper, mag/arcsec$^2$\\
}
  
\longtab{3}{
\begin{longtable}{lllllllll}
\caption{Catalogue of projected fraction-of-light radii (in arcsec) obtained by using the model-independent method presented in Paper I.\label{FoLRadiiTable}}\\
\hline\hline
ID & $\log{r_{10\%}}$ & $\log{r_{20\%}}$ & $\log{r_{30\%}}$ & $\log{r_{40\%}}$  & $\log{r_{60\%}}$ & $\log{r_{70\%}}$ & $\log{r_{80\%}}$ & $\log{r_{90\%}}$ \\
\hline
\endfirsthead
\caption{continued.}\\
\hline\hline
ID & $\log{r_{10\%}}$ & $\log{r_{20\%}}$ & $\log{r_{30\%}}$ & $\log{r_{40\%}}$ & $\log{r_{60\%}}$ & $\log{r_{70\%}}$ & $\log{r_{80\%}}$ & $\log{r_{90\%}}$ \\
\hline
\endhead
\hline\hline
\endfoot
ARP 2        & $1.607$ & $1.777$ & $1.897$ & $1.997$  & $2.197$ & $2.307$ & $2.437$ & $2.607$ \\
FORNAX 1     & $0.785$ & $0.965$ & $1.075$ & $1.155$  & $1.315$ & $1.385$ & $1.475$ & $1.595$ \\
FORNAX 2     & $0.574$ & $0.754$ & $0.874$ & $0.974$  & $1.164$ & $1.274$ & $1.404$ & $1.614$ \\
FORNAX 3     & $0.250$ & $0.430$ & $0.580$ & $0.710$  & $0.980$ & $1.130$ & $1.310$ & $1.560$ \\
FORNAX 4     & $0.213$ & $0.383$ & $0.523$ & $0.653$  & $0.913$ & $1.073$ & $1.263$ & $1.553$ \\
FORNAX 5     & $0.207$ & $0.387$ & $0.547$ & $0.687$  & $0.967$ & $1.117$ & $1.317$ & $1.587$ \\
IC 1276      & $1.663$ & $1.863$ & $2.003$ & $2.133$  & $2.383$ & $2.543$ & $2.773$ & $3.263$ \\
IC 4499      & $1.489$ & $1.699$ & $1.829$ & $1.939$  & $2.139$ & $2.229$ & $2.339$ & $2.489$ \\
LMC-HODGE 11 & $0.920$ & $1.120$ & $1.260$ & $1.380$  & $1.590$ & $1.700$ & $1.810$ & $1.930$ \\
LMC-HODGE 14 & $0.749$ & $0.959$ & $1.109$ & $1.249$  & $1.519$ & $1.669$ & $1.819$ & $2.039$ \\
LMC-NGC 1466 & $0.760$ & $0.960$ & $1.090$ & $1.190$  & $1.400$ & $1.510$ & $1.650$ & $1.850$ \\
LMC-NGC 1711 & $0.790$ & $0.970$ & $1.110$ & $1.230$  & $1.500$ & $1.670$ & $1.860$ & $2.100$ \\
LMC-NGC 1754 & $0.553$ & $0.733$ & $0.893$ & $1.043$  & $1.363$ & $1.543$ & $1.763$ & $2.073$ \\
LMC-NGC 1777 & $0.950$ & $1.150$ & $1.300$ & $1.410$  & $1.640$ & $1.750$ & $1.900$ & $2.090$ \\
LMC-NGC 1786 & $0.704$ & $0.864$ & $1.004$ & $1.134$  & $1.424$ & $1.644$ & $1.954$ & $2.424$ \\
LMC-NGC 1805 & $0.579$ & $0.779$ & $0.929$ & $1.059$  & $1.359$ & $1.569$ & $1.809$ & $2.109$ \\
LMC-NGC 1818 & $0.800$ & $1.000$ & $1.140$ & $1.260$  & $1.500$ & $1.650$ & $1.830$ & $2.130$ \\
LMC-NGC 1831 & $1.011$ & $1.191$ & $1.311$ & $1.421$  & $1.631$ & $1.741$ & $1.881$ & $2.151$ \\
LMC-NGC 1835 & $0.448$ & $0.648$ & $0.788$ & $0.908$  & $1.138$ & $1.278$ & $1.528$ & $2.028$ \\
LMC-NGC 1847 & $0.979$ & $1.239$ & $1.439$ & $1.599$  & $1.859$ & $1.989$ & $2.119$ & $2.239$ \\
LMC-NGC 1850 & $0.940$ & $1.170$ & $1.320$ & $1.460$  & $1.740$ & $1.890$ & $2.050$ & $2.210$ \\
LMC-NGC 1856 & $0.818$ & $1.048$ & $1.218$ & $1.368$  & $1.638$ & $1.778$ & $1.928$ & $2.128$ \\
LMC-NGC 1868 & $0.609$ & $0.809$ & $0.939$ & $1.059$  & $1.299$ & $1.439$ & $1.639$ & $2.009$ \\
LMC-NGC 1898 & $0.830$ & $1.030$ & $1.180$ & $1.310$  & $1.580$ & $1.720$ & $1.840$ & $1.980$ \\
LMC-NGC 1916 & $0.442$ & $0.602$ & $0.732$ & $0.852$  & $1.092$ & $1.232$ & $1.412$ & $1.662$ \\
LMC-NGC 2004 & $0.719$ & $0.909$ & $1.059$ & $1.189$  & $1.489$ & $1.659$ & $1.869$ & $2.109$ \\
LMC-NGC 2005 & $0.356$ & $0.586$ & $0.756$ & $0.916$  & $1.236$ & $1.436$ & $1.696$ & $2.056$ \\
LMC-NGC 2011 & $0.680$ & $0.930$ & $1.130$ & $1.280$  & $1.550$ & $1.670$ & $1.790$ & $1.920$ \\
LMC-NGC 2019 & $0.396$ & $0.616$ & $0.786$ & $0.936$  & $1.276$ & $1.506$ & $1.776$ & $2.066$ \\
LMC-NGC 2031 & $0.950$ & $1.160$ & $1.320$ & $1.470$  & $1.720$ & $1.840$ & $1.970$ & $2.160$ \\
LMC-NGC 2100 & $0.608$ & $0.838$ & $1.008$ & $1.158$  & $1.458$ & $1.638$ & $1.858$ & $2.108$ \\
LMC-NGC 2121 & $1.209$ & $1.379$ & $1.499$ & $1.589$  & $1.739$ & $1.819$ & $1.909$ & $2.029$ \\
LMC-NGC 2136 & $0.630$ & $0.820$ & $0.950$ & $1.060$  & $1.280$ & $1.410$ & $1.620$ & $2.010$ \\
LMC-NGC 2153 & $0.386$ & $0.606$ & $0.756$ & $0.886$  & $1.126$ & $1.246$ & $1.396$ & $1.576$ \\
LMC-NGC 2155 & $1.020$ & $1.230$ & $1.370$ & $1.510$  & $1.760$ & $1.900$ & $2.040$ & $2.180$ \\
LMC-NGC 2157 & $0.830$ & $1.020$ & $1.150$ & $1.270$  & $1.610$ & $1.880$ & $2.260$ & $2.520$ \\
LMC-NGC 2159 & $0.850$ & $1.040$ & $1.190$ & $1.330$  & $1.730$ & $1.990$ & $2.330$ & $2.530$ \\
LMC-NGC 2162 & $0.760$ & $0.960$ & $1.100$ & $1.210$  & $1.420$ & $1.520$ & $1.640$ & $1.780$ \\
LMC-NGC 2172 & $0.890$ & $1.080$ & $1.220$ & $1.360$  & $1.710$ & $1.950$ & $2.180$ & $2.400$ \\
LMC-NGC 2173 & $0.960$ & $1.180$ & $1.340$ & $1.490$  & $1.760$ & $1.910$ & $2.110$ & $2.380$ \\
LMC-NGC 2193 & $0.699$ & $0.929$ & $1.089$ & $1.239$  & $1.539$ & $1.719$ & $1.919$ & $2.149$ \\
LMC-NGC 2210 & $0.670$ & $0.850$ & $0.970$ & $1.070$  & $1.270$ & $1.390$ & $1.550$ & $1.800$ \\
LMC-NGC 2213 & $0.709$ & $0.919$ & $1.069$ & $1.199$  & $1.499$ & $1.689$ & $1.939$ & $2.249$ \\
LMC-NGC 2214 & $0.880$ & $1.110$ & $1.280$ & $1.430$  & $1.760$ & $1.930$ & $2.140$ & $2.480$ \\
LMC-NGC 2231 & $0.901$ & $1.131$ & $1.271$ & $1.381$  & $1.571$ & $1.651$ & $1.751$ & $1.871$ \\
LMC-NGC 2249 & $0.650$ & $0.850$ & $0.990$ & $1.110$  & $1.320$ & $1.440$ & $1.590$ & $1.800$ \\
LMC-NGC 2257 & $1.173$ & $1.363$ & $1.493$ & $1.603$  & $1.803$ & $1.893$ & $1.983$ & $2.073$ \\
LMC-SL 842   & $0.780$ & $0.950$ & $1.070$ & $1.160$  & $1.330$ & $1.430$ & $1.540$ & $1.700$ \\
NGC 104      & $1.440$ & $1.720$ & $1.920$ & $2.090$  & $2.380$ & $2.540$ & $2.690$ & $2.890$ \\
NGC 1261     & $1.100$ & $1.300$ & $1.430$ & $1.530$  & $1.710$ & $1.810$ & $1.940$ & $2.130$ \\
NGC 1851     & $0.731$ & $1.001$ & $1.201$ & $1.351$  & $1.681$ & $1.871$ & $2.141$ & $2.551$ \\
NGC 1904     & $0.967$ & $1.197$ & $1.357$ & $1.497$  & $1.747$ & $1.887$ & $2.057$ & $2.327$ \\
NGC 2419     & $0.860$ & $1.110$ & $1.280$ & $1.410$  & $1.650$ & $1.770$ & $1.920$ & $2.130$ \\
NGC 2808     & $1.070$ & $1.280$ & $1.420$ & $1.540$  & $1.750$ & $1.860$ & $2.000$ & $2.250$ \\
NGC 288      & $1.654$ & $1.854$ & $1.964$ & $2.054$  & $2.204$ & $2.294$ & $2.384$ & $2.504$ \\
NGC 3201     & $1.638$ & $1.818$ & $1.958$ & $2.068$  & $2.278$ & $2.388$ & $2.528$ & $2.718$ \\
NGC 362      & $0.970$ & $1.200$ & $1.350$ & $1.490$  & $1.760$ & $1.920$ & $2.090$ & $2.340$ \\
NGC 4147     & $0.743$ & $1.033$ & $1.193$ & $1.333$  & $1.573$ & $1.713$ & $1.873$ & $2.093$ \\
NGC 4590     & $1.374$ & $1.574$ & $1.704$ & $1.824$  & $2.044$ & $2.174$ & $2.324$ & $2.594$ \\
NGC 5024     & $1.235$ & $1.445$ & $1.595$ & $1.725$  & $1.955$ & $2.095$ & $2.275$ & $2.495$ \\
NGC 5053     & $1.795$ & $1.965$ & $2.065$ & $2.145$  & $2.285$ & $2.355$ & $2.445$ & $2.575$ \\
NGC 5139     & $1.948$ & $2.138$ & $2.258$ & $2.358$  & $2.538$ & $2.638$ & $2.748$ & $2.888$ \\
NGC 5272     & $1.289$ & $1.479$ & $1.619$ & $1.739$  & $1.969$ & $2.089$ & $2.229$ & $2.479$ \\
NGC 5286     & $1.100$ & $1.320$ & $1.470$ & $1.600$  & $1.850$ & $1.990$ & $2.180$ & $2.460$ \\
NGC 5466     & $1.678$ & $1.828$ & $1.938$ & $2.028$  & $2.198$ & $2.278$ & $2.378$ & $2.528$ \\
NGC 5634     & $0.857$ & $1.127$ & $1.307$ & $1.457$  & $1.757$ & $1.917$ & $2.107$ & $2.417$ \\
NGC 5694     & $0.618$ & $0.878$ & $1.068$ & $1.228$  & $1.528$ & $1.708$ & $1.918$ & $2.218$ \\
NGC 5824     & $0.613$ & $0.873$ & $1.073$ & $1.253$  & $1.633$ & $1.853$ & $2.093$ & $2.443$ \\
NGC 5897     & $1.650$ & $1.820$ & $1.930$ & $2.030$  & $2.200$ & $2.280$ & $2.380$ & $2.540$ \\
NGC 5904     & $1.359$ & $1.589$ & $1.759$ & $1.909$  & $2.169$ & $2.289$ & $2.419$ & $2.599$ \\
NGC 5986     & $1.238$ & $1.418$ & $1.538$ & $1.638$  & $1.828$ & $1.928$ & $2.048$ & $2.228$ \\
NGC 6093     & $0.898$ & $1.128$ & $1.308$ & $1.448$  & $1.708$ & $1.828$ & $1.968$ & $2.178$ \\
NGC 6121     & $1.805$ & $2.065$ & $2.235$ & $2.365$  & $2.595$ & $2.725$ & $2.885$ & $3.165$ \\
NGC 6139     & $1.020$ & $1.310$ & $1.520$ & $1.700$  & $2.050$ & $2.260$ & $2.500$ & $2.810$ \\
NGC 6171     & $1.434$ & $1.664$ & $1.834$ & $1.974$  & $2.254$ & $2.414$ & $2.624$ & $2.924$ \\
NGC 6205     & $1.449$ & $1.649$ & $1.779$ & $1.889$  & $2.089$ & $2.199$ & $2.309$ & $2.479$ \\
NGC 6218     & $1.494$ & $1.704$ & $1.844$ & $1.944$  & $2.134$ & $2.234$ & $2.344$ & $2.494$ \\
NGC 6229     & $0.757$ & $0.967$ & $1.117$ & $1.237$  & $1.467$ & $1.577$ & $1.717$ & $1.907$ \\
NGC 6254     & $1.484$ & $1.674$ & $1.814$ & $1.934$  & $2.164$ & $2.284$ & $2.414$ & $2.594$ \\
NGC 6256     & $1.094$ & $1.334$ & $1.484$ & $1.594$  & $1.774$ & $1.854$ & $1.924$ & $2.004$ \\
NGC 6266     & $1.130$ & $1.370$ & $1.530$ & $1.670$  & $1.940$ & $2.090$ & $2.270$ & $2.560$ \\
NGC 6273     & $1.328$ & $1.548$ & $1.698$ & $1.828$  & $2.098$ & $2.258$ & $2.468$ & $2.798$ \\
NGC 6284     & $0.988$ & $1.238$ & $1.418$ & $1.568$  & $1.838$ & $2.008$ & $2.178$ & $2.358$ \\
NGC 6287     & $1.138$ & $1.348$ & $1.488$ & $1.598$  & $1.828$ & $1.978$ & $2.148$ & $2.398$ \\
NGC 6293     & $1.098$ & $1.388$ & $1.588$ & $1.758$  & $2.088$ & $2.278$ & $2.478$ & $2.698$ \\
NGC 6316     & $0.988$ & $1.238$ & $1.428$ & $1.588$  & $1.868$ & $2.018$ & $2.238$ & $2.698$ \\
NGC 6325     & $1.048$ & $1.278$ & $1.468$ & $1.608$  & $1.878$ & $2.048$ & $2.248$ & $2.528$ \\
NGC 6333     & $1.227$ & $1.417$ & $1.547$ & $1.657$  & $1.877$ & $2.007$ & $2.167$ & $2.417$ \\
NGC 6341     & $1.129$ & $1.349$ & $1.499$ & $1.639$  & $1.869$ & $1.999$ & $2.139$ & $2.319$ \\
NGC 6342     & $0.937$ & $1.197$ & $1.387$ & $1.527$  & $1.807$ & $1.927$ & $2.057$ & $2.267$ \\
NGC 6356     & $1.117$ & $1.337$ & $1.507$ & $1.647$  & $1.927$ & $2.097$ & $2.327$ & $2.737$ \\
NGC 6362     & $1.608$ & $1.788$ & $1.918$ & $2.038$  & $2.308$ & $2.488$ & $2.698$ & $2.968$ \\
NGC 6366     & $1.802$ & $1.982$ & $2.102$ & $2.202$  & $2.362$ & $2.452$ & $2.562$ & $2.742$ \\
NGC 6388     & $0.870$ & $1.100$ & $1.250$ & $1.390$  & $1.640$ & $1.770$ & $1.930$ & $2.150$ \\
NGC 6397     & $1.580$ & $1.800$ & $1.960$ & $2.110$  & $2.380$ & $2.520$ & $2.690$ & $2.940$ \\
NGC 6401     & $1.148$ & $1.388$ & $1.568$ & $1.708$  & $1.948$ & $2.068$ & $2.198$ & $2.338$ \\
NGC 6402     & $1.409$ & $1.589$ & $1.709$ & $1.819$  & $1.999$ & $2.099$ & $2.219$ & $2.399$ \\
NGC 6440     & $0.834$ & $1.074$ & $1.234$ & $1.374$  & $1.654$ & $1.814$ & $2.024$ & $2.324$ \\
NGC 6441     & $0.882$ & $1.112$ & $1.292$ & $1.452$  & $1.782$ & $1.972$ & $2.212$ & $2.572$ \\
NGC 6517     & $0.827$ & $1.107$ & $1.297$ & $1.447$  & $1.717$ & $1.867$ & $2.057$ & $2.337$ \\
NGC 6522     & $0.950$ & $1.200$ & $1.380$ & $1.510$  & $1.740$ & $1.840$ & $1.930$ & $2.020$ \\
NGC 6528     & $0.798$ & $1.008$ & $1.168$ & $1.298$  & $1.568$ & $1.728$ & $1.958$ & $2.298$ \\
NGC 6539     & $1.389$ & $1.609$ & $1.769$ & $1.909$  & $2.219$ & $2.419$ & $2.659$ & $2.969$ \\
NGC 6541     & $1.120$ & $1.360$ & $1.540$ & $1.690$  & $1.990$ & $2.160$ & $2.380$ & $2.740$ \\
NGC 6553     & $1.328$ & $1.528$ & $1.658$ & $1.768$  & $1.988$ & $2.108$ & $2.288$ & $2.598$ \\
NGC 6569     & $1.158$ & $1.368$ & $1.508$ & $1.638$  & $1.888$ & $2.038$ & $2.238$ & $2.638$ \\
NGC 6584     & $1.130$ & $1.340$ & $1.480$ & $1.600$  & $1.820$ & $1.930$ & $2.060$ & $2.210$ \\
NGC 6624     & $0.968$ & $1.258$ & $1.458$ & $1.628$  & $1.948$ & $2.128$ & $2.308$ & $2.498$ \\
NGC 6626     & $1.208$ & $1.468$ & $1.648$ & $1.808$  & $2.128$ & $2.328$ & $2.598$ & $3.048$ \\
NGC 6637     & $1.160$ & $1.380$ & $1.530$ & $1.670$  & $1.940$ & $2.100$ & $2.300$ & $2.590$ \\
NGC 6638     & $0.937$ & $1.137$ & $1.277$ & $1.407$  & $1.657$ & $1.797$ & $1.977$ & $2.227$ \\
NGC 6642     & $0.858$ & $1.138$ & $1.338$ & $1.488$  & $1.748$ & $1.878$ & $1.988$ & $2.108$ \\
NGC 6652     & $0.820$ & $1.080$ & $1.260$ & $1.400$  & $1.660$ & $1.810$ & $2.000$ & $2.270$ \\
NGC 6656     & $1.758$ & $1.968$ & $2.118$ & $2.258$  & $2.518$ & $2.678$ & $2.858$ & $3.128$ \\
NGC 6681     & $0.983$ & $1.273$ & $1.463$ & $1.623$  & $1.943$ & $2.123$ & $2.343$ & $2.733$ \\
NGC 6712     & $1.384$ & $1.564$ & $1.704$ & $1.824$  & $2.034$ & $2.114$ & $2.214$ & $2.354$ \\
NGC 6715     & $0.790$ & $1.030$ & $1.210$ & $1.370$  & $1.690$ & $1.890$ & $2.150$ & $2.550$ \\
NGC 6723     & $1.410$ & $1.600$ & $1.730$ & $1.840$  & $2.060$ & $2.160$ & $2.280$ & $2.450$ \\
NGC 6752     & $1.360$ & $1.620$ & $1.800$ & $1.950$  & $2.210$ & $2.360$ & $2.520$ & $2.740$ \\
NGC 6779     & $1.249$ & $1.439$ & $1.579$ & $1.689$  & $1.909$ & $1.989$ & $2.079$ & $2.209$ \\
NGC 6809     & $1.752$ & $1.922$ & $2.032$ & $2.132$  & $2.302$ & $2.392$ & $2.492$ & $2.612$ \\
NGC 6864     & $0.710$ & $0.960$ & $1.140$ & $1.300$  & $1.580$ & $1.720$ & $1.880$ & $2.080$ \\
NGC 6934     & $0.959$ & $1.159$ & $1.309$ & $1.439$  & $1.679$ & $1.809$ & $1.959$ & $2.179$ \\
NGC 6981     & $1.169$ & $1.359$ & $1.499$ & $1.609$  & $1.809$ & $1.919$ & $2.049$ & $2.249$ \\
NGC 7006     & $0.839$ & $1.029$ & $1.169$ & $1.289$  & $1.529$ & $1.679$ & $1.879$ & $2.229$ \\
NGC 7078     & $1.048$ & $1.318$ & $1.508$ & $1.658$  & $1.938$ & $2.088$ & $2.258$ & $2.498$ \\
NGC 7089     & $1.178$ & $1.388$ & $1.528$ & $1.658$  & $1.898$ & $2.028$ & $2.178$ & $2.388$ \\
NGC 7099     & $1.070$ & $1.350$ & $1.530$ & $1.670$  & $1.930$ & $2.070$ & $2.240$ & $2.500$ \\
NGC 7492     & $1.418$ & $1.568$ & $1.668$ & $1.758$  & $1.908$ & $1.978$ & $2.058$ & $2.188$ \\
PAL 10       & $1.557$ & $1.807$ & $1.987$ & $2.137$  & $2.397$ & $2.507$ & $2.617$ & $2.727$ \\
PAL 11       & $1.511$ & $1.691$ & $1.801$ & $1.891$  & $2.041$ & $2.121$ & $2.201$ & $2.331$ \\
PAL 12       & $1.278$ & $1.498$ & $1.648$ & $1.768$  & $1.978$ & $2.098$ & $2.248$ & $2.488$ \\
PAL 13       & $0.893$ & $1.043$ & $1.153$ & $1.243$  & $1.393$ & $1.463$ & $1.543$ & $1.633$ \\
PAL 1        & $1.104$ & $1.254$ & $1.394$ & $1.504$  & $1.694$ & $1.774$ & $1.854$ & $1.954$ \\
PAL 5        & $1.771$ & $1.941$ & $2.051$ & $2.141$  & $2.271$ & $2.341$ & $2.421$ & $2.521$ \\
PAL 8        & $1.408$ & $1.628$ & $1.778$ & $1.908$  & $2.128$ & $2.228$ & $2.338$ & $2.438$ \\
SMC-KRON 3   & $1.072$ & $1.262$ & $1.392$ & $1.492$  & $1.692$ & $1.782$ & $1.892$ & $2.042$ \\
SMC-NGC 121  & $0.800$ & $0.990$ & $1.120$ & $1.250$  & $1.520$ & $1.690$ & $1.950$ & $2.370$ \\
SMC-NGC 152  & $1.161$ & $1.361$ & $1.491$ & $1.601$  & $1.801$ & $1.901$ & $2.021$ & $2.201$ \\
SMC-NGC 176  & $0.780$ & $0.970$ & $1.100$ & $1.220$  & $1.480$ & $1.650$ & $1.830$ & $2.010$ \\
SMC-NGC 361  & $1.073$ & $1.263$ & $1.383$ & $1.473$  & $1.633$ & $1.703$ & $1.783$ & $1.883$ \\
SMC-NGC 411  & $0.820$ & $1.020$ & $1.170$ & $1.300$  & $1.540$ & $1.690$ & $1.870$ & $2.140$ \\
SMC-NGC 416  & $0.720$ & $0.910$ & $1.030$ & $1.140$  & $1.340$ & $1.460$ & $1.610$ & $1.810$ \\
SMC-NGC 458  & $0.860$ & $1.060$ & $1.200$ & $1.330$  & $1.570$ & $1.740$ & $1.990$ & $2.430$ \\
TERZAN 5     & $1.099$ & $1.359$ & $1.539$ & $1.699$  & $1.999$ & $2.159$ & $2.349$ & $2.559$ \\
\end{longtable}
}
\end{document}